\documentclass[twocolumn,showpacs,preprintnumbers,amsmath,amssymb,prb]{revtex4-2}
\usepackage{soul}
\usepackage[bbgreekl]{mathbbol}
\usepackage{graphicx}% Include figure files
\usepackage{dcolumn}% Align table columns on decimal point
\usepackage{bm}% bold math
\usepackage{mathtools}
\usepackage{hyperref}
\usepackage{color}
\usepackage{amsbsy}
\graphicspath{{fig/}{img/}}

\newcommand{\qn}{{\bf  q}}
\newcommand{\pn}{{\bf  p}}
\newcommand{\Kn}{{\bf  K}}

\newcommand{\Rn}{{\bf R}}

\newcommand{\rn}{{\bf r}}

\newcommand{\An}{{\bf A}}

\def\gsim{\lower.35em\hbox{$\stackrel{\textstyle>}{\textstyle\sim}$}}
\def\lsim{\lower.35em\hbox{$\stackrel{\textstyle<}{\textstyle\sim}$}}

\begin{document}
\title{Many-Body Quantum Geometric Dipole}
\author{H.A.Fertig$^{1,2,3}$ and Luis Brey$^3$}
\affiliation{$^1$ Department of Physics, Indiana University, Bloomington, IN 47405}
\affiliation{$^2$ Quantum Science and Engineering Center, Indiana University, Bloomington, IN 47408}
\affiliation{ $^3$ Instituto de Ciencia de Materiales de Madrid (CSIC), Cantoblanco, 28049 Madrid, Spain}

\date{\today}
\begin{abstract}
Collective excitations of many-body electron systems can carry internal structure, tied to the quantum geometry of the Hilbert space in which they are embedded.  This has been shown explicitly for particle-hole-like excitations, which carry a ``quantum geometric dipole'' (QGD) that is essentially an electric dipole moment associated with the state.  We demonstrate in this work that this property can be formulated in a generic way, which does not require wavefunctions expressed in terms of single particle-hole states.  Our formulation exploits the density matrix associated with a branch of excitations that evolves continuously with its momentum $\Kn$, from which one may extract single-particle states allowing a construction of the QGD.  We demonstrate the formulation using the single-mode approximation for excited states of two quantum Hall systems: the first for an integrally filled Landau level, and the second for a fractional quantum Hall state at filling factor $\nu=1/m$, with $m$ an odd integer.  In both cases we obtain the same result for the QGD, which can be attributed to the translational invariance assumed of the system.   Our study demonstrates that the QGD is an intrinsic property of collective modes which is valid beyond approximations one might make for their wavefunctions.
\end{abstract}
\maketitle
\section{Introduction}
\label{sec:introduction}
In recent years the relevance of quantum geometry to condensed matter physics has become increasingly appreciated.  Its realization via Berry phases of single particle states can be used to understand the quantized Hall effect \cite{Thouless_1982,Avron_2003,Girvin-Book}, as well as related states with quantized linear response \cite{Haldane_1988,Kane_2005,Bernevig_2006,Qi_2010,Nagaosa_2010}. More generally, quantum geometry makes its presence felt in single-particle transport in a variety of electron systems
\cite{Sundaram_1999,Haldane_2004,Xiao_2010,Lapa_2019,Torma_2023}.  Non-linear transport can also be understood within quantum geometric frameworks \cite{Sodemann_2015,Ortix_2021,Nagaosa_2024,Xu_2024,Wang_2024}.  There are also important impacts of the quantum geometry of single-particle states on collective ground states, most prominently ones involving superconductivity or superfluidity \cite{Peotta_2015,Julku_2016,Liang_2017,Iskin_2018,Xie_2020,Rossi_2021,Verma_2021,Ahn_2021,Simon_2022,Herzog_2022,Huhtinen_2022,Torma_2022,Peotta_2023,Hu_2023,Tian_2023,Mao_2023,Verma_2023,Uchoa_2023}.

The quantum geometry of single-particle states can also impact properties associated with collective modes in a system.  For example, for insulating bands with non-trivial Berry curvature, the spectrum of exciton energies deviate from the usual simple hydrogenic form \cite{Srivastava_2015,Zhou_2015}.  In metallic systems, the single-particle geometry  impacts plasmon dynamics \cite{Song_2016,Arora_2022}.  For such collective modes, however, quantum geometry impacts the system in a more direct way.  In particular, for systems with translational symmetry, collective modes carry a momentum quantum number, and the continuous evolution of the excited states with this momentum label allows other possibilities to characterize the geometry and topology of the Hilbert space in which they are embedded \cite{Yao_2008,Kwan_2021}.
For example, excitons -- bound states of a particle and hole, usually in a band structure -- carry an intrinsic Berry curvature, distinct from those of the single-particle bands hosting the constituents \cite{Yao_2008,Kwan_2021}.

The fact that collective modes host internal structure without analog in single-particle states suggests that they may have unique quantum geometric properties.  Recently this has been demonstrated for neutral excitations that can be described by particle-hole pairs.  Specifically, one finds an intrinsic dipole moment that can be understood in terms of the collective mode wavefunction evolution with its momentum.  This \textit{quantum geometric dipole} (QGD) is a natural property of excitons \cite{Cao_PRB_2021,Tang_2023}, and a direct consequence for two-dimensional systems is that an applied electric field $\pmb{\mathcal{E}}$ induces exciton drift perpendicular to that field, analogous to the $\pmb{\mathcal{E}} \times {\bf B}$ drift one expects for charged particles in a magnetic field ${\bf B}$ \cite{Cao_PRB_2021,Chaudhary_2021}.  An analogous QGD may be present for two-dimensional plasmons, which can be described as particle-hole excitations across a Fermi surface \cite{Sawada:1957aa}.  The QGD for these excitations results in an asymmetry of scattering from circularly symmetric scattering potentials that is present only when the QGD is non-vanishing \cite{Cao_PRL_2021}.  In quasi-one-dimensional systems, plasmons may also carry a transverse dipole moment which is closely related to the corresponding two-dimensional QGD in the large wire width limit \cite{Cao_2022}.  Interestingly, such transverse dipole moments can also appear in excitons of quasi-one-dimensional insulators \cite{Ulloa_2002}.

\begin{widetext}
\begin{figure*}[th]
	\centering
	\includegraphics[width=\textwidth]{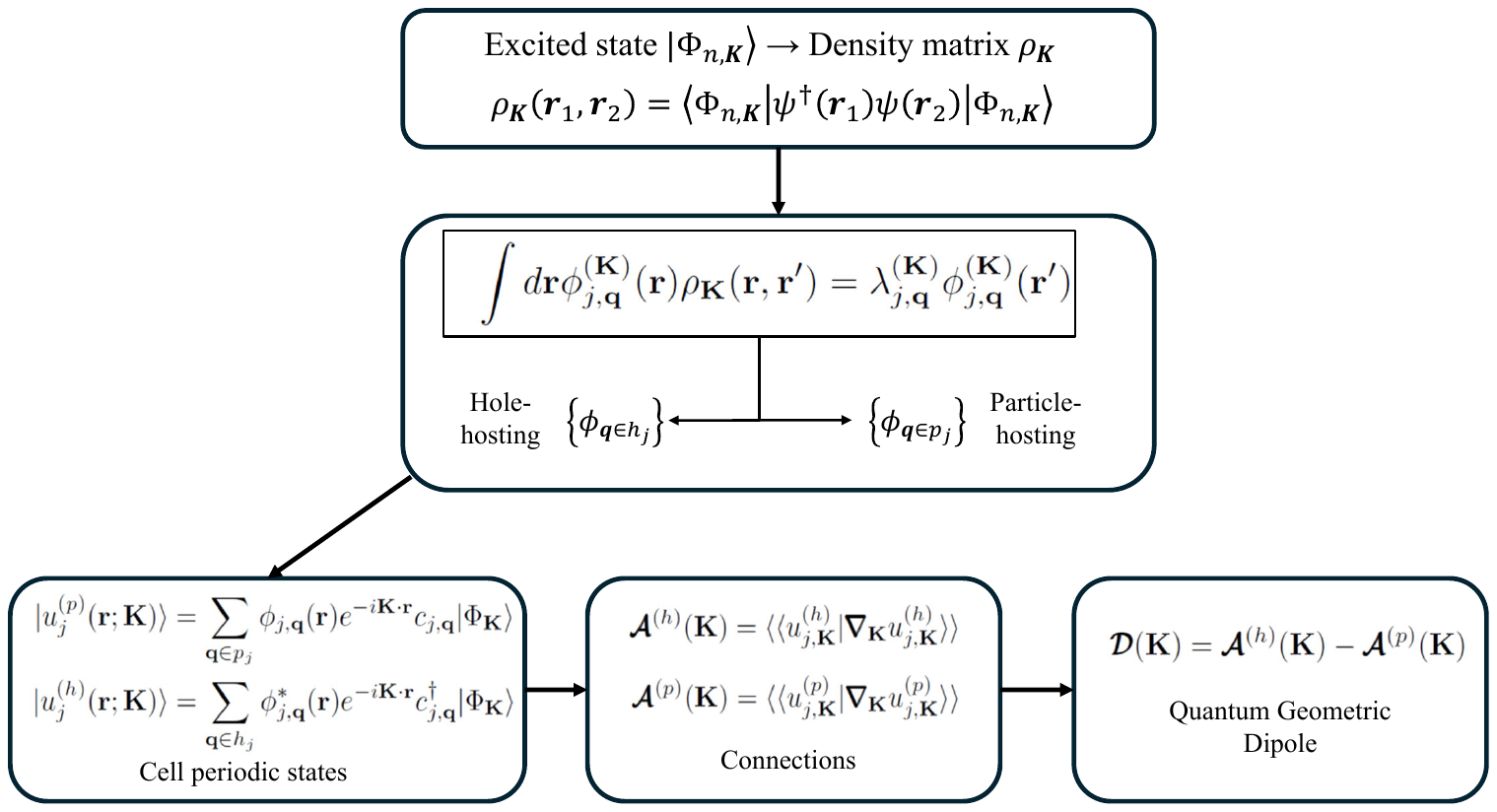}
	\caption{Basic steps followed to find the QGD for a band $n$ of many-body states $\{ \Phi_{n,{\bf K}} \}$ with momenta ${\bf K}$.  One begins by computing a $\Kn$-dependent density matrix $\rho_{\bf K}$, whose single-particle eigenstates $\phi_{j,{\bf q}}$ are divided into two groups, hole-hosting ($h_j$) and particle-hosting ($p_j$).  Cell-periodic functions of each type are then constructed for each band $j$, and from these one can define connections associated with holes ($\pmb{\mathcal{A}}^{(h)}(\Kn)$) and with particle ($\pmb{\mathcal{A}}^{(p)}(\Kn)$).  Their discrete difference defines the quantum geometric dipole $\pmb{\mathcal{D}}(\Kn)$.  Details of each step are discussed in the main text.}
\label{fig:FlowDiagram}
\end{figure*}
\end{widetext}

The two-body forms for wavefunctions of these types of excitations are convenient, and in many cases accurate, approximations to their true many-body state.  Nevertheless, in almost all cases exact wavefunctions will contain corrections that involve states with more than just a single particle-hole pair.
Moreover, some systems, particularly highly correlated ones, may possess excitations with well-defined momenta that are not well-described in terms of an effective two-body state.  This raises the question of whether the QGD is a well-defined concept which can be applied to more general wavefunction forms.  In what follows, we demonstrate that indeed it is.
We introduce a formalism allowing a computation of the quantum geometric dipole of a many-body state with momentum ${\bf K}$ of {\it generic} form.  Briefly, the formalism involves using the density operator associated with the system state for each ${\bf K}$ to define a set of single-particle states.  These states may be divided into two groups, one which we call {\it particle-hosting}, and the other {\it hole-hosting}, with some flexibility in precisely how states are assigned to each group.  With these two sets of states defined, one then defines connections $\mathcal{A}^{(p)}({\bf K})$ and $\mathcal{A}^{(h)}({\bf K})$ associated with particles and holes, respectively, which are analogs of the Berry connection for single-particle band states \cite{Vanderbilt_book}.  Their discrete difference represents the quantum geometric dipole $\pmb{\mathcal{D}}({\bf K})$, and we will show this has the natural interpretation of an internal dipole moment for the many-body state.  Fig. \ref{fig:FlowDiagram} summarizes the basic steps we use to define the quantum geometric dipole.
\newpage
To demonstrate the validity of the approach, we consider two explicit examples, in each case computing the QGD of an excitation mode above a quantum Hall state \cite{Girvin-Book,Yoshioka_book}.  These may be found in two-dimensional electron gas systems immersed in a perpendicular magnetic field $B\hat{z}$.  The first of our examples is the low-lying magnetoexciton mode above an integrally filled Landau level.  In the strong field limit, this is well-described by states involving a single electron in an otherwise empty Landau level, bound to a single hole in an otherwise filled Landau level \cite{Lerner_1979,Bychkov_1981,Kallin_1984}.  In this limit, a mode of momentum $\hbar {\bf K}$ carries an intrinsic dipole moment ${\bf p} = e {\bf K} \times \hat{z} \ell^2$, where $\ell=\sqrt{\hbar c/eB}$, with $c$ the speed of light and $e$ the charge of an electron \cite{Lerner_1979,Bychkov_1981,Kallin_1984}.  This has been shown to be consistent with the QGD of this neutral mode, for which one finds $\pmb{\mathcal{D}} = {\bf K} \times \hat{z} \ell^2$ \cite{Cao_PRB_2021}.  Moreover, this value of $\pmb{\mathcal{D}}$ turns out to be precisely what is needed for the exciton equations of motion to be effectively Lorentz invariant \cite{Cao_PRB_2021}.  For the present study, we adopt a different form for the magneto-exciton wavefunctions, specifically one generated using the single mode approximation (SMA) \cite{Feynman-Book,Girvin_1985,Girvin_1986}.  In this case the state is not restricted to a single pair of Landau levels, and so is not limited to strong fields, but it does fall into the paradigm of (a linear combination of) single particle-hole pair states.  We find that our many-body approach to the QGD produces exactly the same result as expected in the strong field case ($\pmb{\mathcal{D}} = {\bf K} \times \hat{z} \ell^2$), as should be expected since the effective Lorentz invariance in this problem does not require a strong magnetic field.

Our second example involves magnetoplasmon excitations above a partially filled lowest Landau level in a strong magnetic field, with filling factor (defined as the ratio of electron density to magnetic flux density in the electron gas) $\nu=1/m$, with $m$ an odd integer.  For $m=3$ and $m=5$ such systems are well-known to support the fractional quantum Hall effect (FQHE) \cite{Tsui_1982,Girvin-Book,Yoshioka_book}.  Their ground states are qualitatively well-described by Laughlin-Jastrow wavefunctions \cite{Laughlin_1983,Haldane_1985,Girvin-Book,Yoshioka_book}, with charged excitations of $\pm e/m$. The states may be qualitatively understood in terms of composite fermion theory \cite{Jain_book}, in which a singular gauge transformation attaches flux quanta to electrons (yielding ``composite fermions''), such that at the mean-field level, {the ratio of particle density to magnetic flux density is an integer.}
The state may then be understood in terms of integrally filled Landau levels of composite fermions \cite{Jain_book}.  This suggests a connection between the magnetoplasmons of the fractionally filled system and the magnetoexcitons of the integrally filled one.  With the reduced magnetic flux, the effective magnetic length $\ell^*$ satisfies $\ell^{*2}=m\ell^2$.  These two observations suggest some tension between the reduced charge of the quasiparticles relative to the integer case, which presumably lowers the dipole moment of a collective mode, and the smaller effective magnetic field, which tends to raise it.

The resolution of this tension can be found by direct calculation of the QGD, $\pmb{\mathcal{D}}$.  To do this we use
approximate wavefunctions for the magnetoplasmons, generated using the SMA \cite{Girvin_1985,Girvin_1986}.  The computation of $\pmb{\mathcal{D}}$ for these states is more involved than in the integral case, but, as we show below, it may be carried through without further approximation beyond the wavefunctions themselves.  The final result is remarkably simple, and is in fact identical to the result for integrally filled Landau levels.  This shows that the two effects discussed above essentially cancel against one another.

While this precise cancellation is at first surprising, it is in fact necessary that it should happen. We show this by demonstrating that {\it any} state with well-defined momentum $\hbar {\bf K}$ which lies fully in the lowest Landau must have an electric dipole moment perpendicular to that momentum, with magnitude $eK\ell^2$, where $\ell$ is the magnetic length associated with the physical magnetic field.  Thus, our many-body approach to the QGD produces a correct result for the FQHE example.  This suggests that our formulation indeed gives physically sensible results for states that cannot be described within a simple single particle-hole paradigm.  More generally, we see an internal structure associated with the state which can be understood quantum geometrically, without a priori assumptions that its wavefunction has a particular form.

This paper is organized as follows.  In Section II, we explain in detail our method for computing the QGD of a many-body state of generic form.  Section III is devoted to our example of the magnetoexciton excitation above a filled Landau level.  In Section IV, we present our analysis of the QGD for a fractional quantum Hall state at filling factor $\nu=1/m$ with $m$ an odd integer, using excited states generated by the single mode approximation.  While the details of this turn out to be involved, the final result is quite simple.  We show why this is the case in Section V.  Section VI summarizes our study, and discusses possible future directions for this work.  Our paper also has two appendices.  Appendix A presents a study of a possible offset term that appears when connecting the quantum geometric dipole to the physical electric dipole moment.  We argue that this is essentially a constant term for low energy excitations, which in many interesting cases vanishes.  Appendix B presents some details of the QGD calculation for the magnetoexciton excitation above a filled Landau level.

\section{Many-Body Quantum Geometric Dipole}
Our formal development of the quantum geometric dipole (QGD) assumes the Hamiltonian commutes with some set of translation operators $T_{{\bf a}_i}$, $i=1,...D$, where $D$ is the dimensionality of the system, ${\bf a}_i$ are  primitive lattice vectors, and $[T_{{\bf a}_i},T_{{\bf a}_j}]=0$.  In this situation eigenstates of the Hamiltonian $|\Phi_{n,\bf K}\rangle$ may be labeled by a momentum ${\bf K}$ (we herein set $\hbar=1$), for which $T_{{\bf a}_i} |\Phi_{n,\bf K}\rangle =   e^{i{\bf K} \cdot {\bf a}_i} |\Phi_{n,\bf K}\rangle$.  We assume $\Phi_n({\bf K})$, varies continuously with ${\bf K}$, and the index $n$ provides any further quantum numbers needed to specify the state. The states are normalized as $\langle \Phi_{n,{\bf K}}|\Phi_{n',\bf {K}'}\rangle = \delta({\bf K} - {\bf K}') \delta_{n,n'}$.  For simplicity, in this work we focus on Hamiltonians and eigenstates of spinless fermions.

\begin{figure}[th]
    \centering
	\includegraphics[width=1.3\linewidth,trim = 50 150 0 100 ,clip]{"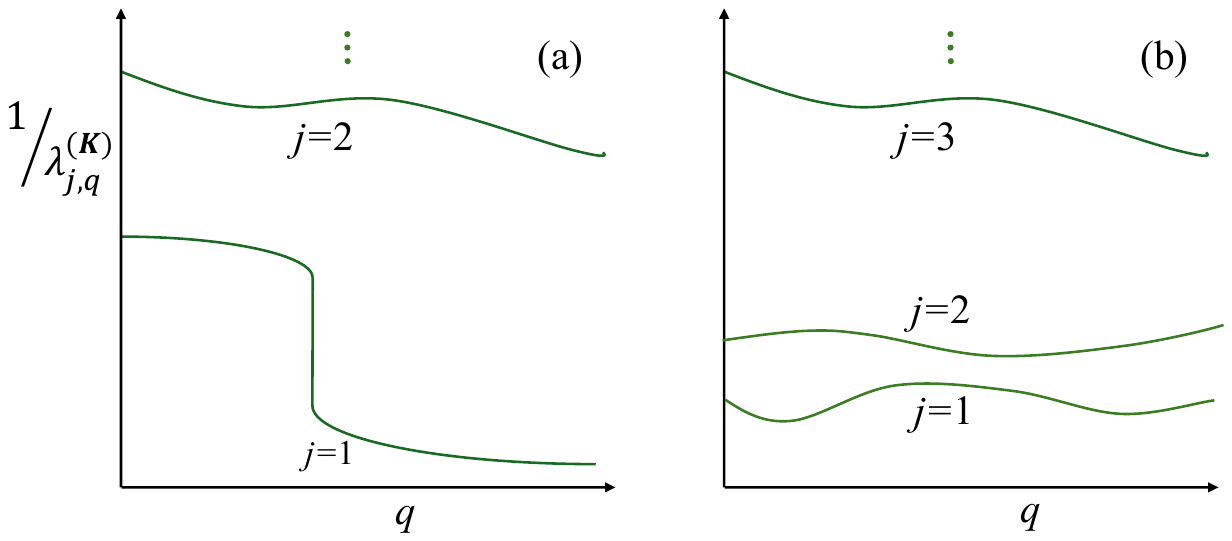"}
	\caption{Qualitative inverse occupations $1/\lambda_{j,{\bf q}}$ of single particle states diagonalizing the density matrix $\rho_{\Kn}$.  (a) Expected form for metal with a single occupied band.  Discontinuity in $\lambda_{j,{\bf q}}$ indicates a Fermi surface.  (b) Expected form for an insulator with a multiple occupied bands.}
\label{fig:Occupations}
\end{figure}

For fixed $n$, from these states we can form as set of density matrices $\rho_{\bf K}({\bf r},{\bf r}') = \langle \Phi_{n,{\bf K}}|\psi^{\dag}({\bf r}) \psi({\bf r}') |\Phi_{n,\bf {K}}\rangle$, where $\psi({\bf r})$ is an annihilation field operator.  (Since we will always work within a fixed $n$ subspace in what follows, for ease of notation we hereon suppress the index $n$.) The density matrices have eigenvalues and eigenfunctions
$$
\int d {\bf r}  \phi^{({\bf K})}_{j,{\bf q}}({\bf r}) \rho_{\bf K}({\bf r},{\bf r}')  = \lambda_{j,{\bf q}}^{({\bf K})} \phi^{({\bf K})}_{j,{\bf q}}({\bf r}').
$$
In writing this, we have noted that the translation operators $T_{{\bf a}_i}$ commute with $\rho_{\bf K}$, so that the eigenfunctions and eigenvalues of the latter can be labeled by a wavevector ${\bf q}$, which in general must be contained within the Brillouin zone associated with the primitive lattice vectors.  The index $j$ labels different discrete eigenstates of $\rho_{\bf K}$ with fixed ${\bf q}$.

The eigenstates $\phi_{j,{\bf q}}^{({\bf K})}$ are single-particle in nature, and their associated
eigenvalues $\lambda_{j,{\bf q}}^{({\bf K})}$ may be viewed as an effective occupation: $\sum_j \sum_{\bf q} \lambda_{j,{\bf q}}^{({\bf K})} |\phi_{j,{\bf q}}^{({\bf K})}({\bf r})|^2$ precisely reproduces the fermion density of the state $|\Phi_{\bf K}\rangle$, $\rho_{\bf K}({\bf r},{\bf r})$.  Fig. \ref{fig:Occupations} illustrates possible expected qualitative behaviors for $\lambda_{j,{\bf q}}^{({\bf K})}.$   We note that as a function of ${\bf q}$, the eigenvalues $\lambda_{j,{\bf q}}$, and the associated states $|\phi_{j,{\bf q}}^{({\bf K})}\rangle$ are organized into bands, analogous to but distinct from energy bands of a single-particle Hamiltonian.

The key to our formulation of the many-body QGD is a division of the states $\{ \phi^{({\bf K})}_{j,{\bf q}} \}$ into two groups, one of which we call ``hole-hosting states,'' and the other ``particle-hosting states.''    In situations where the state $|\Phi_{\bf K}\rangle$ is a Slater determinant, the natural choice for the former would be states for which $\lambda_{j,{\bf q}}^{({\bf K})}=1$, while the latter would be those for which $\lambda_{j,{\bf q}}^{({\bf K})}=0$.   More generally, we will require that the division should be carried out in such a way that gradients of $\{ \phi^{({\bf K})}_{j,{\bf q}} \}$ with respect to {\bf K} are well-defined, and such that the total number of hole-hosting states should be equal to the number of fermions in the system.  Beyond this,
there is some freedom in choosing which group a given $\phi^{({\bf K})}_{j,{\bf q}}$ should be placed in, as in principle this does not affect the final result for the QGD.  However, when approximations are introduced, some choices will work better than others.  Presumably, the best results will be obtained by assigning states with the largest values of $\lambda_{j,{\bf q}}^{({\bf K})}$ to the hole-hosting states.

To define the QGD in this setting, we construct ${\bf K}$-dependent spinor states,
\begin{align}
|u_{j}({{\bf r};\bf K})\rangle =
\begin{pmatrix}
|u_j^{(p)}({\bf r};{\bf K})\rangle \\
|u_j^{(h)}({\bf r};{\bf K})\rangle
\end{pmatrix}
\label{eqn:spinor_def}
\end{align}
where
\begin{align}
|u_j^{(p)}({\bf r};{\bf K})\rangle &= \sum_{{\bf q} \in p_j} \phi_{j,{\bf q}}({\bf r}) e^{-i{\bf K} \cdot {\bf r}} c_{j,{\bf q}} |\Phi_{\bf K}\rangle, \label{eqn:particle_def} \\
|u_j^{(h)}({\bf r};{\bf K})\rangle &= \sum_{{\bf q} \in h_j} \phi^{*}_{j,{\bf q}}({\bf r}) e^{-i{\bf K} \cdot {\bf r}} c^{\dag}_{j,{\bf q}} |\Phi_{\bf K}\rangle.
\label{eqn:hole_def}
\end{align}
In these expressions, $\sum_{{\bf q} \in p_j}$ denotes a sum over particle-hosting states in band $j$, $\sum_{{\bf q} \in h_j}$ denotes a sum over hole-hosting states in band $j$, and $c^{({\bf K})}_{j,{\bf q}}$ annihilates a particle of momentum ${\bf q}$ in band $j$.  Note that the construction of $|u_j^{(p)}({\bf r};{\bf K})\rangle$ gives it non-zero contributions from particles residing within the manifold of particle-hosting states, so that $\sum_j\langle u_j^{(p)}({\bf r};{\bf K})|u_j^{(p)}({\bf r};{\bf K})\rangle$ may be interpreted as a density of particles at position ${\bf r}$, while $\sum_j\langle u_j^{(h)}({\bf r};{\bf K})|u_j^{(h)}({\bf r};{\bf K})\rangle$ is the corresponding hole density.  Having constructed these states, we endow them with two further properties.  The first is their behavior under translations, which we define to be
\begin{align}
\tilde{T}_{\bf a}|u_j^{(p)}({\bf r};{\bf K})\rangle &= \sum_{{\bf q} \in p_j} \phi^{({\bf K})}_{j,{\bf q}}({\bf r}+{\bf a}) e^{-i{\bf K} \cdot ({\bf r}+{\bf a})} T_{\bf a} \, c^{({\bf K})}_{j,{\bf q}} |\Phi_{\bf K}\rangle, \nonumber \\
\tilde{T}_{\bf a}|u_j^{(h)}({\bf r};{\bf K})\rangle &= \sum_{{\bf q} \in h_j} \phi^{({\bf K})*}_{j,{\bf q}}({\bf r}+{\bf a}) e^{-i{\bf K} \cdot ({\bf r}+{\bf a})} T_{\bf a} \, c^{({\bf K})\dag}_{j,{\bf q}} |\Phi_{\bf K}\rangle.
\nonumber
\end{align}
{Note that these operators act on all the particles represented by the state $|\Phi_{\Kn}\rangle$, as well as on the parameter $\rn$.}
It is not difficult to show $|u_j^{(p,h)}({\bf r};{\bf K})\rangle$ is invariant under $\tilde{T}_{{\bf a}_i}$, where ${\bf a}_i$ is a primitive lattice vector.  This is important in that it means we are putting vectors at different ${\bf K}$ into a single vector space (i.e., Eqs. \ref{eqn:particle_def} and \ref{eqn:hole_def} represent affine connections for the states \cite{Coleman_book}), so that taking derivatives with respect to ${\bf K}$ becomes meaningful.
The second is an inner product, which we define as
\begin{equation}
\label{eqn:inner_product}
\langle\langle u_{j,{\bf K}} | u_{j',{\bf K}'} \rangle\rangle
\equiv \int d{\bf r}
\langle u_{j}({{\bf r};\bf K}) | u_{j'}({{\bf r};\bf K}') \rangle.
\end{equation}
The integration over ${\bf r}$ in this expression is over all of real space.

The quantum geometric quantity of interest to us is
\begin{align}
\pmb{\mathcal{D}}({\bf K}) &= i\sum_j \left[ \langle\langle u_{j,{\bf K}}^{(h)} | \pmb{\nabla}_{\bf K} u_{j,{\bf K}}^{(h)} \rangle\rangle - \langle\langle u_{j,{\bf K}}^{(p)} | \pmb{\nabla}_{\bf K} u_{j,{\bf K}}^{(p)} \rangle\rangle \right] \nonumber \\
&\equiv \pmb{\mathcal{A}}^{(h)}({\bf K}) - \pmb{\mathcal{A}}^{(p)}({\bf K}).
\label{eqn:QGD}
\end{align}
From its form, we see that $\pmb{\mathcal{D}}({\bf K})$ is a measure of how the states $| u_{j}({{\bf r};\bf K}) \rangle$ change as one moves through the parameter space ${\bf K}$, and in this sense it is a quantum geometric quantity.  Physically it is essentially an electric dipole moment associated with the many-body state $|\Phi_{\bf K}\rangle$, in units where the fermion charge is taken as unity.  To see this, define
\begin{widetext}
\begin{align}
|\psi_{j,{\bf K}}({\bf r})\rangle &\equiv
\begin{pmatrix}
|\psi_{j,{\bf K}}^{(p)}({\bf r})\rangle \\
|\psi_{j,{\bf K}}^{(h)}({\bf r})\rangle
\end{pmatrix}
%\nonumber \\
%&
=
e^{i{\bf K} \cdot {\bf r}}
\begin{pmatrix}
|u_{j}^{(p)}({{\bf r};\bf K})\rangle \\
|u_{j}^{(h)}({{\bf r};\bf K})\rangle
\end{pmatrix}, \nonumber
\end{align}
so that
%\begin{widetext}
\begin{equation}
\pmb{\mathcal{D}}({\bf K})
=
\sum_j\int d{\bf r}\left\{
%\left[ e^{i{\bf K} \cdot {\bf r}} \pmb{\nabla}_{\bf K} e^{-i{\bf K} \cdot {\bf r}} \right]
{\bf r} \left[
\langle \psi_{j,{\bf K}}^{(h)}({\bf r})|\psi_{j,{\bf K}}^{(h)}({\bf r}) \rangle -
\langle \psi_{j,{\bf K}}^{(p)}({\bf r})|\psi_{j,{\bf K}}^{(p)}({\bf r}) \rangle
\right]
-i\langle \psi_{j,{\bf K}}({\bf r})|\sigma_z|\pmb{\nabla}_{\bf K}\psi_{j,{\bf K}}({\bf r}) \rangle
\right\},
\label{eqn:showing_dipole}
\end{equation}
where $\sigma_z$ is a Pauli matrix.
This expression is well-defined provided $N^{(p)}_{\bf K}-N^{(h)}_{\bf K}=0$, where
\begin{align}
N^{(p)}_{\bf K} &= \sum_j \langle\langle \psi_{j,{\bf K}}^{(p)}({\bf r})|\psi_{j,{\bf K}}^{(p)}({\bf r}) \rangle\rangle, \nonumber \\
N^{(h)}_{\bf K} &= \sum_j \langle\langle \psi_{j,{\bf K}}^{(h)}({\bf r})|\psi_{j,{\bf K}}^{(h)}({\bf r}) \rangle\rangle. \nonumber
\end{align}
\end{widetext}
In particular the first integral on the right-hand side of Eq. \ref{eqn:showing_dipole} is independent of the choice of origin of ${\bf r}$, and the second term, which can be rewritten as
$$
\lim_{{\bf K}' \to {\bf K}} \pmb{\nabla}_{\bf K} \left[ \left( N^{(p)}_{\bf K} - N^{(h)}_{\bf K} \right) \delta ({\bf K}- {\bf K}' ) \right],
$$
vanishes.  Because $\pmb{\mathcal{D}}({\bf K})$ is independent of the origin of coordinates, it is clear that it represents information about the $\textit{internal}$ structure of the excited state.

{The quantities $N^{(p)}_{\bf K}$ and $N^{(h)}_{\bf K}$ have the interpretations of the average number of particles in the particle-hosting states, and the average number of holes in the hole-hosting states, respectively.}
The requirement $N^{(p)}_{\bf K}-N^{(h)}_{\bf K}=0$ can be imposed in a simple way.  It proceeds by noticing that the state $|\Phi_{\bf K}\rangle$ can be constructed starting from a Slater determinant in which all the hole-hosting states are filled, and all the particle-hosting states are empty.  The excited state $|\Phi_{\bf K}\rangle$ is in principle a linear combination of states in which different numbers of fermions are excited out of the hole-hosting states into the particle-hosting states. {Because the excitation of a particle out of the initial Slater determinant necessarily leaves a hole behind,} every term in this linear combination {\it individually} has equal numbers of particle and holes.  Then the averages $N^{(p)}_{\bf K}$ and $N^{(h)}_{\bf K}$ must be equal.

Two comments are in order.  First, from the above discussion, one sees that if the total number of hole-hosting states is equal to the total number of fermions in the system, we will always meet the condition $N^{(p)}_{\bf K}-N^{(h)}_{\bf K}=0$.  In what follows, we enforce $N^{(p)}_{\bf K} = N^{(h)}_{\bf K}$  by adopting this condition on the number of hole-hosting states.  Second, this construction explains why the precise choice of particle- and hole-hosting states does not affect $\pmb{\mathcal{D}}({\bf K})$: interchanging a pair of states between the two groups changes $\langle \psi_{j,{\bf K}}^{(p)}({\bf r})|\psi_{j,{\bf K}}^{(p)}({\bf r}) \rangle$ and $\langle \psi_{j,{\bf K}}^{(h)}({\bf r})|\psi_{j,{\bf K}}^{(h)}({\bf r}) \rangle$ by precisely equal amounts, and $N^{(p)}_{\bf K}-N^{(h)}_{\bf K}$ still vanishes.

To complete the demonstration that $\pmb{\mathcal{D}}({\bf K})$ represents a dipole moment, we connect its form back to the density matrix.  With some algebra, one may show
\begin{widetext}
\begin{align}
\pmb{\mathcal{D}}({\bf K}) &=
\sum_j\int d{\bf r} \,
%\left[ e^{i{\bf K} \cdot {\bf r}} \pmb{\nabla}_{\bf K} e^{-i{\bf K} \cdot {\bf r}} \right]
{\bf r} \left[
\langle \psi_{j,{\bf K}}^{(h)}({\bf r})|\psi_{j,{\bf K}}^{(h)}({\bf r}) \rangle -
\langle \psi_{j,{\bf K}}^{(p)}({\bf r})|\psi_{j,{\bf K}}^{(p)}({\bf r}) \rangle
\right] \nonumber \\%
%&= \sum_j \left\{
%\sum_{{\bf q}_1,{\bf q}_2 \in h_j}
%\int d^Dr \,\phi_{j,{\bf q}_1}^{({\bf K})}({\bf r})  \, {\bf r} \, \phi_{j,{\bf q}_2}^{({\bf K})*}({\bf r})
%\langle \Phi_{\bf K} | c_{j,{\bf q}_1} c_{j,{\bf q}_2}^{\dag} | \Phi_{\bf K} \rangle
%-
%\sum_{{\bf q}_1,{\bf q}_2 \in p_j}
%\int d^Dr \,\phi_{j,{\bf q}_1}^{({\bf K})*}({\bf r}) \, {\bf r} \, \phi_{j,{\bf q}_2}^{({\bf K})}({\bf r})
%\langle \Phi_{\bf K} | c_{j,{\bf q}_1}^{\dag} c_{j,{\bf q}_2} | \Phi_{\bf K} \rangle
%\right\}
%\nonumber \\
%&= \sum_j \left\{
%\sum_{{\bf q} \in h_j} \int d^Dr \, {\bf r} \, |\phi_{j,{\bf q}}^{({\bf K})}({\bf r})|^2
%-\sum_{{\bf q}} \int d^Dr \, {\bf r} \, |\phi_{j,{\bf q}}^{({\bf K})}({\bf r})|^2
%\langle \Phi_{\bf K} | c_{j,{\bf q}}^{\dag} c_{j,{\bf q}} | \Phi_{\bf K} \rangle
%\right\} \nonumber \\
%&= \sum_j \left\{
%\sum_{{\bf q} \in h_j} \int d^Dr \, {\bf r} \, |\phi_{j,{\bf q}}^{({\bf K})}({\bf r})|^2
%-\sum_{{\bf q}} \int d^Dr \, {\bf r} \, |\phi_{j,{\bf q}}^{({\bf K})}({\bf r})|^2 \lambda_{j,{\bf q}}
%\right\}\nonumber \\
&=
\left\{\sum_j\sum_{{\bf q} \in h_j} \int d{\bf r} \, {\bf r} \, |\phi_{j,{\bf q}}^{({\bf K})}({\bf r})|^2 \right\}
-{\rm Tr} \left( {\bf r}\rho_{\bf K} \right) \equiv {\bf R}_0^{({\bf K})} -{\rm Tr} \left( {\bf r}\rho_{\bf K} \right).
\label{R0_def}
\end{align}
\end{widetext}
The last term is the dipole moment of the charge density associated with $|\Phi_{\bf K}\rangle$.  The term ${\bf R}_0^{({\bf K})}$ is the dipole moment of the Slater determinant state formed by placing exactly one fermion in each hole-hosting state.  In general, this means $\pmb{\mathcal{D}}({\bf K})$ actually represents the \textit{deviation} of the dipole moment from that of a reference state.

{Two further comments are in order.  First, both the physical dipole moment and that of the reference state depend on the origin of coordinates.  However, the QGD itself is independent of this choice, as expected for a quantity that characterizes the internal structure of the states \cite{comment}.  Second, the reference state dipole moment, ${\bf R}_0^{({\bf K})}$,
may itself be ${\bf K}$-dependent.  However, in practical situations, it is not.  Moreover, if we assume that the set of hole-hosting states at ${\bf K}=0$, $\{ \phi^{(0)}_{j,{\bf q}} \}$, does not pick some direction in space (as should be possible, for example, of a system with inversion symmetry), then it is natural to have ${\bf R}_0^{({\bf K})}=0$.  We discuss the behavior of ${\bf R}_0^{({\bf K})}$ in more detail in Appendix A.}

Eq. \ref{eqn:QGD} is our definition of the quantum geometric dipole, formulated in a such a way that it can be constructed for a general many-body state, without any specific assumptions about its form.  To test its validity, we now consider two concrete examples, both for excited states of quantum Hall systems in two dimensions.  Our first is the QGD of a magnetoexciton, above a filled Landau level.

\section{QGD for Magnetoexcitons Above a Filled Landau Level}

For two-dimensional electrons in a perpendicular magnetic field, the non-interacting energy spectrum breaks up into Landau levels with energies $\hbar\omega_c(n+{1 \over 2})$, where $\omega_c=eB/mc$ is the cyclotron frequency, $B$ is the magnetic field, $m$ the electron mass, and $n$ is a non-negative integer.  Each of these Landau levels is highly degenerate, with the number of states in a Landau level (LL) being equal to the number of magnetic flux quanta through the two-dimensional system.  One way to label the states in a Landau level
is to imagine the electrons being subject to an infinitesimal spatially periodic potential, with a single magnetic flux quantum through each unit cell.  In this case the Landau level states can be written as eigenstates of the magnetic translation group \cite{Girvin-Book,Cao_PRB_2021}. Each state in a Landau level is then labeled by a unique wavevector.

\begin{figure*}[th]
    \centering
	\includegraphics[width=\linewidth,trim = 0 150 0 200 ,clip]{"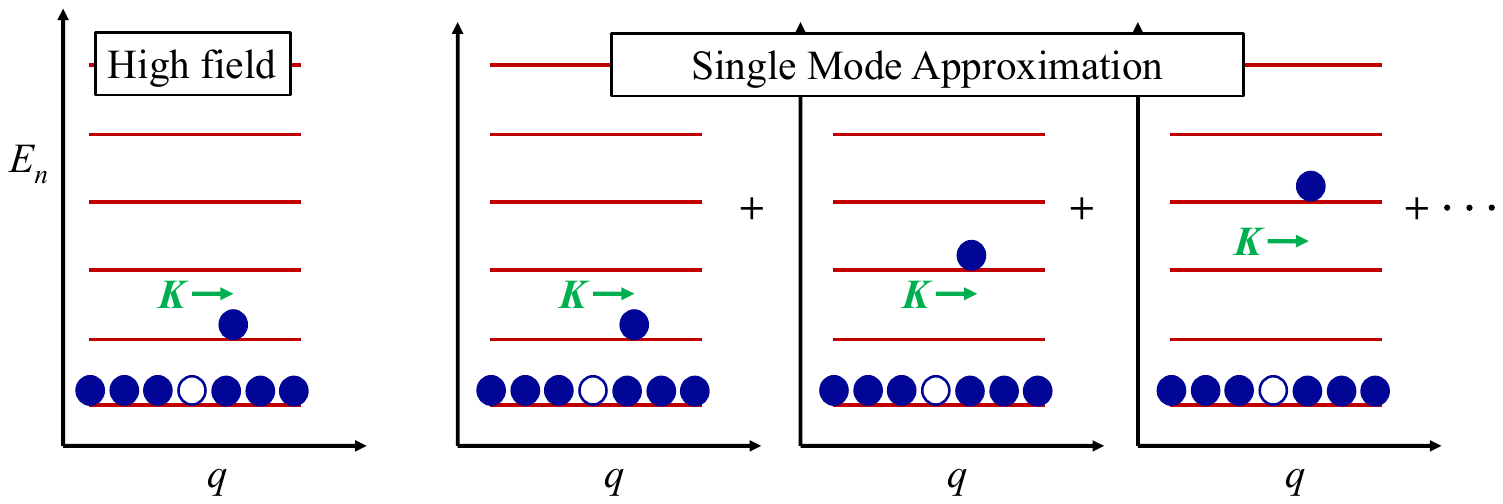"}
	\caption{Qualitative distinction between strong field wavefunction for magnetoexciton above a filled Landau level (left), and its wavefunction in the single mode approximation (SMA) (right).  In the former, a particle is excited from the filled level into the lowest lying empty Landau level, and is given a boost of momentum $\Kn$.  In the SMA, the particle is excited into a linear combination of unoccupied Landau levels, but is given the same momentum boost.  Note that the actual wavefunctions involve linear combinations of states with different wavevectors for the hole, $\qn$ (not shown.)}
\label{fig:SMA}
\end{figure*}

The electron density  $\rho_{el}$ of such a system is parameterized by the filling factor $\nu=2\pi\ell^2\rho_{el}$, where $\ell=\sqrt{\hbar c/eB}$ is the magnetic length.  When $\nu$ is integral, the number of electrons is sufficient to fully fill $\nu$ Landau levels.  In the strong field limit, such a state forms a good approximation to the ground state.  Low-energy neutral excitations can be created by exciting an electron out of the highest occupied Landau level, into the lowest unoccupied Landau level.  By considering linear combinations of such particle-hole states in which each has a fixed momentum difference between the particle and hole, one constructs a wavefunction for the excitation with well-defined momentum \cite{Cao_PRB_2021}. (See Fig. \ref{fig:SMA}.)  Because the state involves creating a bound state particle-hole pair across a single-particle energy gap, this excitation can be understood as a magnetoexciton.

An alternative method for constructing these excitations, which is less dependent on taking the strong field limit, is the single mode approximation (SMA) \cite{Girvin_1985,Girvin_1986}.  This involves acting on the ground state with the density operator
$$Q_{\bf K}^{\dag} \equiv \sum_{i=1}^N e^{i{\bf K} \cdot {\bf r}_i},$$
where ${\bf r}_i$ are the positions of the $N$ particles in the system.  Such a state has momentum ${\bf K}$, but also captures the ground state correlations expected to remain important in a low-energy excitation \cite{Feynman-Book}.  In general one need not assume the ground state has the form of a single filled Landau level to apply the SMA; moreover, because of the form of $Q_{\bf K}^{\dag}$, the particle resides in a linear combination of higher Landau levels (Fig. \ref{fig:SMA}.) One may show that this approximation produces the correct exact excitation energy $\hbar\omega_c$ in the limit $\Kn \to 0$, and also saturates the oscillator-strength sum rule \cite{Girvin_1986}.

As a first example, we apply our approach to computing the QGD to the magnetoexciton of a single filled Landau level ($\nu=1$) as described by the SMA.

\begin{widetext}

\subsection{Density Matrix $\rho_{\Kn}(\rn,\rn')$}

The first step in computing the QGD in this many-body formulation requires we find eigenstates of the $\Kn$-dependent density operator,
\begin{equation}
\rho_{\Kn}(\rn,\rn') = \int d^2r_2 d^2r_3 \dots d^2r_N \Phi_{\Kn}^*(\rn,\rn_2, \dots,\rn_N)\Phi_{\Kn}(\rn',\rn_2, \dots,\rn_N),
\end{equation}
where $\Phi_{\Kn}$ is the excited state wavefunction,
\begin{equation}
\Phi_{\Kn} = {1 \over \sqrt{\mathcal{N}}_{\Kn}} Q_{\Kn}^{\dag} \Psi_0,
\end{equation}
and $\Psi_0$ is the ground state, which we approximate as a filled $n=0$ Landau level. $\mathcal{N}_{\Kn}$ normalizes this state.  Although our formalism allows for the possibility that the eigenstates of $\rho_{\Kn}(\rn,\rn')$ are $\Kn$-dependent, we will see in the thermodynamic limit that there is no such dependence.  The eigenstates turn out to be the single-particle Landau level states.

To see this, we exploit the antisymmetry of $\Phi_0(\rn_1,\rn_2,\dots)$ to write
\begin{align}
\rho_{\Kn}(\rn,\rn') &=  {1 \over \mathcal{N}_\Kn} \int d^2r_2 d^2r_3 \dots d^2r_N \Phi_0^*(\rn,\rn_2,\rn_3,\dots,\rn_N)
               \Phi_0(\rn',\rn_2,\rn_3,\dots,\rn_N) \nonumber \\
           &\times \left\{e^{i\Kn \cdot (\rn-\rn')} + (N-1)\left[  e^{i\Kn \cdot (\rn-\rn_2)}+e^{i\Kn \cdot (\rn_2-\rn')} \right] + N(N-1) e^{i\Kn \cdot (\rn_2-\rn_3)} \right\}.
\label{eqn:PhiStarPhi}
\end{align}
\end{widetext}
In this approximation, $\Phi_0$ is a Slater determinant, consisting of $N!$ terms, each a product of $N$ single-particle states, all with different momentum labels $\qn$, the sum of which should vanish.  The integrations over $\rn_2 \dots \rn_N$ fix which pairs of terms in $\Phi_0^*$ and $\Phi_0$ can yield non-zero values.  For each term  in the second line of Eq. \ref{eqn:PhiStarPhi} there are $N!$ non-vanishing terms.  Thus in the large $N$ limit, the very last term dominates the result.

Because of this, in the thermodynamic limit, the density matrix must have the form
\begin{equation}
\rho_{\Kn}(\rn,\rn') = \sum_{\bf q}
\alpha_{\qn} \phi_{0,\qn}^*({\bf r})\phi_{0,\qn}({\bf r}')
\end{equation}
where $\alpha_{\qn}$ are real numbers, and $\phi_{0,\qn}$ are lowest Landau level states.  It immediately follows that any state of the lowest Landau level is an eigenstate of $\rho_{\Kn}$ with non-zero eigenvalue, and all higher Landau level states are zero eigenvalue states.  Since at $\nu=1$ there are $N$ states in a Landau level, we adopt the lowest Landau level states as our hole-hosting states, and all higher Landau level states as particle-hosting states.

\subsection{Computation of the QGD}
\label{sec:filledLL}
To compute the QGD we need explicit expressions for the states defined in Eqs. \ref{eqn:particle_def} and \ref{eqn:hole_def}.  The quantities $|u_j^{(p)}({\bf r};{\bf K})\rangle$ and $|u_j^{(h)}({\bf r};{\bf K})\rangle$
involve distinct sets of bands (in the case of $|u_j^{(h)}\rangle$ only a single band is involved), but because one sums over {\it all} the states in a band, some simplification possible.  In particular we can choose any form of basis states that fully covers these bands to perform the sums.  In this application, a particularly simple choice involves eigenstates of the non-interacting Hamiltonian $H=({\hbar \over i} \pmb{\nabla} + {e \over c} {\bf A})^2$ with vector potential in the Landau gauge, $\An = Bx\pmb{\hat{y}}$.  The single-particle states are then
\begin{equation}
\phi_{n,X}(\rn) = \frac{1}{\sqrt{\pi^{1/2} 2^n n! \ell L_y}}e^{-iXy/\ell^2}H_n(x-X) e^{-(x-X)^2/2\ell^2},
\nonumber
\end{equation}
with $n$ the Landau level index, $L_y$ the extent of the system along the $\hat{y}$ direction, $H_n$ a Hermite polynomial, and $X$ the guiding center quantum number labeling states within a Landau level.  In terms of these states we have
\begin{align}
|u_n^{(p)}({\bf r};{\bf K})\rangle &= \sum_{X} \phi_{n,X}({\bf r}) e^{-i{\bf K} \cdot {\bf r}} c_{n,X} |\Phi_{\bf K}\rangle \nonumber \\
|u_0^{(h)}({\bf r};{\bf K})\rangle &= \sum_{X} \phi^{*}_{0,X}({\bf r}) e^{-i{\bf K} \cdot {\bf r}} c^{\dag}_{0,X} |\Phi_{\bf K}\rangle
\nonumber
\end{align}
where $c_{n,X}$ annihilates an electron in state $\phi_{n,X}$.  Note that, because the sets of particle- and hole-hosting states do not depend on $\Kn$, we have dropped the $(\Kn)$ superscript on the field operators.

To construct the QGD, we form the quantities
\begin{widetext}
\begin{align}
\Gamma_n(\Kn_1,\Kn_2) \equiv \langle\langle u_{n,\Kn_1} | u_{n,\Kn_2} \rangle\rangle
&= \sum_{X_1,X_2}\langle \phi_{n,X_1} |e^{i(\Kn_1-\Kn_2) \cdot \rn} | \phi_{n,X_2} \rangle
\langle \Phi_{\Kn_1} | c^{\dag}_{n,X_1}c_{n,X_2} | \Phi_{\Kn_2} \rangle, \quad {n} > 0, \label{eqn:Gamma_n}\\
&= \sum_{X_1,X_2}\langle \phi_{0,X_2} |e^{i(\Kn_1-\Kn_2) \cdot \rn} | \phi_{n,X_1} \rangle
\langle \Phi_{\Kn_1} | c_{n,X_1}c_{n,X_2}^{\dag} | \Phi_{\Kn_2} \rangle, \quad {n} = 0.  \label{eqn:Gamma_0}
\end{align}
%\end{widetext}
Note that the first factors on the right-hand side of the above equations are single-particle matrix elements, while the second factors are many-body matrix elements. In terms of these, from Eq. 5 we have
\begin{equation}
\pmb{\mathcal{D}}(\Kn)=-i\lim_{\Kn_2 \to \Kn_1}\pmb{\nabla}_{\Kn_2}\left[\sum_{n>0}\Gamma_n(\Kn_1,\Kn_2) - \Gamma_0(\Kn_1,\Kn_2) \right].
\label{eqn:D}
\end{equation}

%\begin{widetext}
The computations of $\sum_{n>0}\Gamma_n(\Kn_1,\Kn_2)$ and $\Gamma_0(\Kn_1,\Kn_2)$ are somewhat involved but are in principle straightforward.  We present some details for these these in Appendix B.  The results are
\begin{align}
\sum_{n>0} \Gamma_n(\Kn_1,\Kn_2) = &
\frac{g}{\sqrt{\mathcal{N}_{\Kn_1}\mathcal{N}_{\Kn_2}}}\left\{e^{-i\hat{z} \cdot (\Kn_1 \times \Kn_2)\ell^2} - e^{-i\hat{z} \cdot (\Kn_1 \times \Kn_2)\ell^2/2 -K_1^2\ell^2/4 - K_2^2\ell^2/4} \right\}
+ \mathcal{O}(\delta K)^2,
\nonumber \\
\Gamma_0(\Kn_1,\Kn_2) = &\frac{g}{\sqrt{\mathcal{N}_{\Kn_1}\mathcal{N}_{\Kn_2}}}
\left\{
1-e^{i\hat{z} \cdot (\Kn_1 \times \Kn_2)\ell^2/2 -K_1^2\ell^2/4 - K_2^2\ell^2/4}
\right\}
+ \mathcal{O}(\delta K)^2.
\end{align}
where $\delta \Kn \equiv \Kn_1 - \Kn_2$ and $g = L_xL_y/2\pi\ell^2$ is the degeneracy of a Landau level, with $L_xL_y$ the system area.
The normalization factors are given by \cite{Girvin_1986} $\mathcal{N}_{\Kn} = \langle \Phi_0 | Q_{\bf K} Q^{\dag}_{\Kn} | \Phi_0 \rangle = g(1-e^{-K^2\ell^2/2})$.  Finally, from Eq. \ref{eqn:D}, we obtain
\begin{align}
\pmb{\mathcal{D}}(\Kn_1)&=\frac{-i}{1-e^{-K_1^2\ell^2/2}}\lim_{\Kn_2 \to \Kn_1} \pmb{\nabla}_{\Kn_2}
\left\{ e^{-i\hat{z} \cdot (\Kn_1 \times \Kn_2)\ell^2} + 2\sin \left[ (\hat{z} \cdot \Kn_1 \times \Kn_2) \ell^2/2 \right] e^{-K_1^2\ell^2/4-K_2^2\ell^2/4} - 1
\right\} \nonumber \\
&=\Kn_1 \times \hat{z} \ell^2. \label{eqn:D_nu=0}
\end{align}
\end{widetext}
This result is identical to that found previously \cite{Cao_PRB_2021} using a different excited state wavefunction for the magnetoexciton, in which only a single Landau level was retained for the particle excited out of the filled band.  In the strong-field limit, the latter yields a lower excitation energy than the SMA wavefunction we have used here \cite{Lerner_1979,Bychkov_1981,Kallin_1984,Girvin_1986}.  The form of the QGD in this context is important, because an applied electric field $\pmb{\mathcal{E}}$ couples to the electric dipole moment, causing a drift motion with velocity ${\bf v}_D = c (\pmb{\mathcal{E}} \times {\bf B})/B^2$, exactly the velocity at which one must move relative to the lab frame for the electric field to vanish \cite{Cao_PRB_2021}.  This means the system is effectively Lorentz invariant, and that the SMA wavefunction for the magnetoexciton respects this symmetry.

Finally, we note that the magnetoexciton states above $\nu=1$ generated by the SMA represent a linear combination of states with a single particle and a single hole, so that the QGD associated with them could have been computed using the methods of Ref. \onlinecite{Cao_PRB_2021}.  The present analysis shows that our many-body approach also produces sensible results for such states.  We next turn our attention to an example that {\it cannot} be represented in terms of single particle-hole pair states, and so requires the method developed above to compute the QGD: magnetoplasmons above a fractional quantum Hall state.

\section{QGD for Magnetoplasmons above a Laughlin State}

We next consider the QGD for a collective excitation above a fractional quantum Hall state, specifically focusing on  filling factors of the form $\nu=1/m$, with $m$ an odd integer.  In a disk geometry, the unnormalized ground state for such a filling is well-described by a Laughlin wave function,
\begin{equation}
\Psi_0({\bf r}_1,{\bf r}_2,\dots) =  \prod_{i<j} (z_i-z_j)^m \prod_k e^{-|z_k|^2/4\ell^2},
\label{eqn:LaughlinState}
\end{equation}
where $z_i=x_i-iy_i$ is the particle position in complex notation, and we have assumed the vector potential to be in symmetric gauge, $\An = \frac{B}{2}(-y,x,0)$.  This wave function has well-defined total angular momentum, and so
does {\it not} have total momentum as a quantum number. While one may write down an analog of this state for electrons on a torus \cite{Haldane_1985}, allowing for momentum quantum numbers -- a fact which we will exploit below -- Eq. \ref{eqn:LaughlinState} is more convenient for explicit calculations. We will see that in the thermodynamic limit, the absence of translational invariance can be overcome.

It is well-known that the low-energy charged excitations above a Laughlin state carry charge $\pm e/m$ \cite{Laughlin_1983,Yoshioka_book}.  In analogy with the types of excitations discussed for filled Landau levels, it is nature to think of the low-lying neutral excitations as bound pairs of such quasiparticles with opposite charge.
This interpretation suggests that the SMA may work well as an approximation for these states.  An important caveat however is that, in the very strong field limit ($\hbar\omega_c \to \infty$), one expects that higher Landau levels will not be involved in any low-energy excitation. Such excitations then involve motion of electrons within a {\it single} Landau level.  Concretely, the (approximate) wavefunctions for these excitations take the form \cite{Girvin_1985,Girvin_1986}
\begin{equation}
\Psi_{\Kn}(\rn_1,\rn_2,\dots) = \frac{1}{\sqrt{\mathcal{N}_{\Kn}^{(m)}}}\overline{Q}_{\Kn}^{\dag} \Psi_0^{(m)}({\bf r}_1,{\bf r}_2,\dots)
\label{SMA}
\end{equation}
where $\mathcal{N}_{\Kn}^{(m)}$ is chosen to normalize the wave function, and $$\overline{Q}_{\bf K}^{\dag} \equiv \sum_{i=1}^N P_0e^{i{\bf K} \cdot {\bf r}_i}P_0$$ is the same operator as used in the last section, but projected by the operator $P_0$ into the lowest Landau level (LLL).  The effect of acting on the partially filled Landau level with $\overline{Q}_{\Kn}^{\dag}$ is to introduce density-wave-like correlations \cite{Feynman-Book,Girvin_1985,Girvin_1986}, and because of this it is natural to think these excitations as magnetoplasmons.
The states $\Psi_{\Kn}^{(m)}$ have been extensively studied \cite{Girvin_1985,Girvin_1986} and are known to have a number of sensible and attractive properties as low-lying collective excitations.

\begin{widetext}

\subsection{Density Matrix for $\nu=1/m$ Laughlin States and its Eigenstates}

The eigenstates of the density matrices associated with these states can be arrived at following reasoning analogous what we followed for integrally filled Landau levels above.  A first observation is that the form of the density matrix,
$$
\rho_{\bf K}(\rn,\rn') \equiv
\int d^2r_2 d^2r_3 \dots d^2r_N \Psi^{(m)*}_{\Kn}  (\rn,\rn_2,\rn_3,\dots)\Psi^{(m)}_{\Kn}  (\rn',\rn_2,\rn_3,\dots),
$$
has the property $\int d^2r \varphi(\rn) \rho_{\bf K}(\rn,\rn') = 0$ for {\it any} state $\varphi$ that lies outside the lowest Landau level.  All such states are then eigenstates of the density matrix, and naturally form part of the particle-hosting states for this system.  The zero eigenvalue, however, indicates that they will not contribute to the QGD.

Because the number of states in the LLL exceeds the number of electrons in the system, we need to divide these into hole-hosting and particle-hosting groups.  As discussed above, the precise division does not in principle affect the final result, provided all states are eigenstates of $\rho_{\Kn}$, the number of hole-hosting states is the same as the number of electrons in the system, and states within each group vary in such a way that derivatives with respect to ${\bf K}$ are well-defined.  A convenient way to proceed in this case is to consider placing the system on a torus, for which an analog of the Laughlin wave function, Eq. \ref{eqn:LaughlinState}, may be written in terms of elliptic $\theta$-functions \cite{Haldane_1985}.  Moreover, one may construct the low-lying collective excitations using the SMA \cite{Repellin_2014}, in a way directly analogous to that described above for the disk geometry.

The utility of considering the torus geometry is that eigenstates of the Hamiltonian are simultaneously eigenstates of a set of translation operators, such that they have good momentum quantum numbers.  In particular one may consider magnetic translations of the form $T_{a\hat{x}} = \prod_{j=1}^N e^{i(p_{j,x}+y_j/2\ell^2)a}$, $T_{a\hat{y}} = \prod_{j=1}^N e^{i(p_{j,y}-x_j/2\ell^2)a}$, where $p_{j,x} = \frac{\hbar}{i}\partial_{x_j}$, $p_{j,y}=\frac{\hbar}{i}\partial_{y_j}$ are momentum operators for particle $j$, for classifying states in the lowest Landau level.  This imposes an effective square lattice structure of lattice constant $a$, and provided there is one magnetic flux quantum per unit cell (i.e., $a^2/2\pi \ell^2=1$), $T_{a\hat{x}}$ and $T_{a\hat{y}}$ commute with each other as well as with the Hamiltonian.  Imposing periodic boundary conditions such that there are $N_c$ unit cells in the whole system, the eigenstates of the Hamiltonian will have well-defined center-of-mass momentum $\Kn$ with $N_c$ possible values.  In particular we must have $T_{a\hat{x}}\Psi_{\Kn} = e^{iK_xa} \Psi_{\Kn}$, $T_{a\hat{y}}\Psi_{\Kn} = e^{iK_ya} \Psi_{\Kn}$.

It follows that the density matrix, viewed as an operator, is invariant under translations, i.e., $T_{\bf a} \rho_{\Kn}(\rn,\rn') T^{-1}_{\bf a} = \rho_{\Kn}(\rn,\rn')$, where ${\bf a}=a\hat{x}$ or $a\hat{y}$.  Its eigenstates $\phi_{0,{\bf q}}(\rn)$ (here the subscript $0$ refers to the LLL) are themselves eigenstates of translations, $T_{\bf a} \phi_{0,{\bf q}}(\rn) = e^{i\qn \cdot {\bf a}} \phi_{0,{\bf q}}(\rn)$, with the number of distinct values of $\qn$ being the same as the number of states in the LLL.  This means that varying $\Kn$ will not change the eigenstates of the density matrix, and our particle-hosting and hole-hosting state can be chosen in a $\Kn$-independent way.
In the following section, we describe one way in which this can be done that ultimately allows a computation of the QGD.

\subsection{Formal Expression for QGD}

For concrete calculations, it is preferable to work with the wave functions in Eqs. \ref{eqn:LaughlinState} and \ref{SMA} than with their counterparts on the torus.  However, this introduces a difficulty in that, for any finite size system, there is an edge which breaks the translational symmetry.  Well inside the bulk, we expect that, locally, states for a disk and states for torus will be essentially the same.
To take advantage of this, we start with formal steps best defined on the torus, and then carry through concrete calculations for needed correlation functions in the disk geometry, for which the calculations are analytically tractable.  We assume the thermodynamic limit has been taken so that the edge does not contribute to these correlation functions.

To compute the QGD, we start with
{
\begin{align}
\Gamma(\Kn,\Kn') &= \langle\langle u_{0,\Kn}^{(p)} | u_{0,\Kn'}^{(p)} \rangle\rangle - \langle\langle u_{0,\Kn}^{(h)} | u_{0,\Kn'}^{(h)} \rangle\rangle \nonumber \\
&= \sum_{\qn_1,\qn_1' \in p_0}
\langle \phi_{0,\qn_1} | e^{i(\Kn-\Kn')\cdot \rn}
|\phi_{0,\qn_1'} \rangle \langle \Psi_{\Kn} | c_{0,\qn_1}^{\dag}  c_{0,\qn_1'} | \Psi_{\Kn'} \rangle
%\nonumber \\
%&-
-\sum_{\qn_2,\qn_2' \in h_0}
\langle \phi_{0,\qn_2'} | e^{i(\Kn-\Kn')\cdot \rn}
|\phi_{0,\qn_2} \rangle \langle \Psi_{\Kn} | c_{0,\qn_2}  c_{0,\qn_2'}^{\dag} | \Psi_{\Kn'} \rangle.
\label{overlap}
\end{align}
The first and second terms of $\Gamma$ can be used to form the particle and hole connections $\pmb{\mathcal{A}}^{(p)}$ and $\pmb{\mathcal{A}}^{(h)}$, respectively, by taking gradients of these terms with respect to $\Kn$.  We then have
\begin{align}
\pmb{\mathcal{D}}(\Kn) &= -\pmb{\mathcal{A}}^{(p)}(\Kn)+\pmb{\mathcal{A}}^{(h)}(\Kn) =
-i \lim_{\Kn' \to \Kn} \pmb{\nabla}_{\Kn'} \Gamma(\Kn,\Kn').
\label{fuzzy_D}
\end{align}
}
Writing $\delta\Kn \equiv \Kn-\Kn'$ allows us to express this as
\begin{align}
\pmb{\mathcal{D}}(\Kn) &= -i \lim_{\delta\Kn \to 0}  \left[\Gamma(\Kn,\Kn-\delta\Kn) - \Gamma(\Kn,\Kn) \right]/\delta K,
\label{D_as_lim}
\end{align}
and from momentum conservation,
\begin{align}
\Gamma(\Kn,\Kn-\delta\Kn) &=
\sum_{\qn_1 \in p_0}
\langle \phi_{0,\qn_1} | e^{i\delta\Kn \cdot \rn}
|\phi_{0,\qn_1-\delta\Kn} \rangle \langle \Psi_{\Kn} | c_{0,\qn_1}^{\dag}  c_{0,\qn_1-\delta\Kn} | \Psi_{\Kn-\delta\Kn} \rangle
\nonumber \\
%&-
&-\sum_{\qn_2 \in h_0}
\langle \phi_{0,\qn_2 +\delta\Kn} | e^{i\delta\Kn \cdot \rn}
|\phi_{0,\qn_2} \rangle \langle \Psi_{\Kn} | c_{0,\qn_2}  c_{0,\qn_2+\delta\Kn}^{\dag} | \Psi_{\Kn-\delta\Kn} \rangle
\label{delta_K}
\end{align}
\end{widetext}

\begin{figure}[th]
    \centering
	\includegraphics[width=1.8\linewidth,trim = 220 150 0 150 ,clip]{"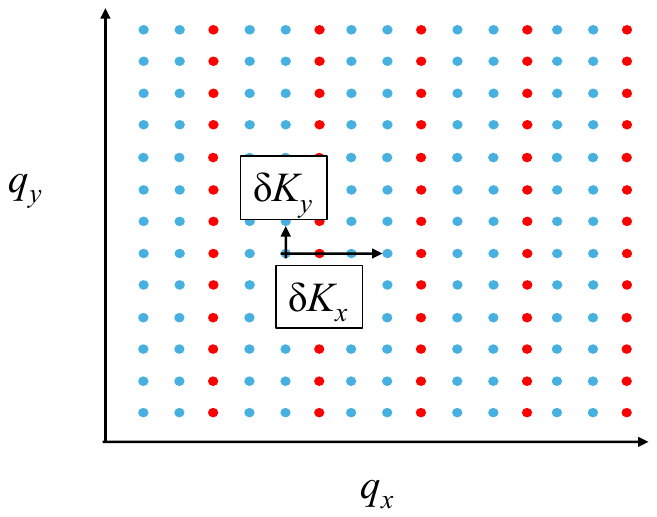"}
	\caption{Illustration of allowed wavevectors for single particle states for system with periodic boundary conditions.  Blue points represent particle-hosting states, red points are hole-hosting.  The lattice and the discrete difference vector $\delta \Kn = (\delta K_x,\delta K_y)$ are constructed so that finite difference approximations to the derivatives of the single particle states with respect to $\delta \Kn$ are always within the particle- or hole-hosting states, so that the limit $\delta \Kn \to 0$, taken in the thermodynamic limit, behaves smoothly.}
\label{fig:Lattice}
\end{figure}

Eqs. \ref{D_as_lim} and \ref{delta_K} raise a subtlety discussed in our initial formulation of the QGD:
since $\pmb{\mathcal{D}}$ is expressed as a limit, we need $\Gamma(\Kn,\Kn-\delta \Kn)$ to behave smoothly as $\delta \Kn \to 0$.  We can in fact guarantee this for both terms in Eq. \ref{delta_K} separately, by a careful division of the states into the particle- and hole-hosting groups, with a corresponding definition of how the $\delta\Kn \to 0$ limit is taken.  To do this, we observe that
if $\qn$ is among the particle-hosting states, then $\qn + \delta \Kn$ must be as well.  Analogous relations should hold for values of $\qn$ among the hole-hosting states.  One way to guarantee this is to group $m$ neighboring values of $\qn$ -- for example, along a row in the $\hat{x}$-direction -- into unit cells, assigning one of these to $h_0$ and the remainder to $p_0$.  The resulting unit cell has length $2\pi/N_c a$ in the $\hat{y}$-direction, and length $2\pi m/N_c a$ in the $\hat{x}$-direction.  (See Fig. \ref{fig:Lattice}.)
We then restrict values of $\Kn$ to discrete points, one associated with each of these unit cells.  This guarantees that in Eq. \ref{delta_K}, matrix elements are always between states in the same sectors. $\delta \Kn$ is then taken to be a discrete difference between the locations of nearby unit cells, and
the limit $\delta \Kn \to 0$ is accomplished by taking the thermodynamic limit, $N_c \to \infty$.

It is interesting to note that this means the number of momenta allowed to the magnetoplasmons above a given ground state will be $1/m$ of the number of unit cells in the system, which in this construction is the number of magnetic flux quanta passing through it.  On a torus, however, there are $m$ different ground states \cite{Haldane_1985,Girvin-Book} upon which one may build a magnetoplasmon in the SMA, so the total number of excited states in a magnetoplasmon band is ultimately equal to the number of flux quanta through the system.  We leave it to future work to determine whether this counting remains valid beyond the SMA.

\begin{widetext}
Continuing with the calculation, we commute two of the fermion operators to get
\begin{align}
\Gamma(\Kn,\Kn') =
\sum_{{\rm all} \, \qn_1}
&\langle \phi_{0,\qn_1} | e^{i\delta\Kn \cdot \rn}
|\phi_{0,\qn_1-\delta\Kn} \rangle \langle \Psi_{\Kn} | c_{0,\qn_1}^{\dag}  c_{0,\qn_1-\delta\Kn} | \Psi_{\Kn'} \rangle
%\nonumber \\
%&-
-N_h\delta(\Kn-\Kn'),
\end{align}
where $N_h$ is the number of hole-hosting states.  Since ultimately we will have $\delta \Kn$ very small, we expand for small $\delta K$.  Noting $\sum_{\qn_1}\langle \Psi_{\Kn} |  c_{0,\qn_1}^{\dag}  c_{0,\qn_1} | \Psi_{\Kn} \rangle = \nu N_c$ is the number of electrons in the system, and that
$$
\sum_{{\rm all} \, \qn_1}
\langle \phi_{0,\qn_1} | e^{i\delta\Kn \cdot \rn}
|\phi_{0,\qn_1-\delta\Kn} \rangle c_{0,\qn_1}^{\dag}  c_{0,\qn_1-\delta\Kn} \equiv \overline{Q}^{\dag}_{\delta \Kn},
$$
we arrive at
\begin{align}
\Gamma(\Kn,\Kn') =&
\sum_{{\rm all} \, \qn_1} \left[
\langle \Psi_{\Kn} | \overline{Q}^{\dag}_{\delta \Kn} | \Psi_{\Kn'} \rangle
-N_h \right] \delta(\Kn-\Kn') \nonumber \\
=&
(\nu N_c-N_h)\delta(\Kn-\Kn') + \lim_{\qn \to 0} \delta \Kn \cdot \pmb{\nabla}_{q}
\langle \Psi_{\Kn} | \overline{Q}^{\dag}_{\qn} | \Psi_{\Kn - \qn} \rangle + \mathcal{O}(\delta K^2).
\end{align}
Our construction requires the first term of the last line to vanish, and the $\mathcal{O}(\delta K^2)$ correction does not contribute to the QGD.  We finally arrive at the expression
\begin{align}
\mathcal{D}(\Kn)
&= -i\lim_{\qn \to 0}
 \pmb{\nabla}_{q}
\langle \Psi_{\Kn} | \overline{Q}^{\dag}_{\qn} | \Psi_{\Kn- \qn} \rangle.
\label{final_D}
\end{align}

In arriving at this expression we have not made any specific assumptions about the form of $|\Psi_{\bf K}\rangle$, except to say that it has a well-defined momentum quantum number, which implies we are working on a torus.  A concrete form of Eq. \ref{final_D}, however, is most easily arrived at by working in the disk geometry.  To compute it we will work in the latter, assuming that the thermodynamic limit has been taken, so that edge effects do not enter.
As we shall see, this can be written, in the SMA, in terms of ground state correlation functions.

To this end, it is convenient to write
\begin{align}
\mathcal{D}(\Kn)
&= -i  \lim_{\qn \to 0}\pmb{\nabla}_{q} \langle \Psi_{\Kn+{1 \over 2}\qn} | \overline{Q}^{\dag}_{\qn} | \Psi_{\Kn- {1 \over 2}\qn} \rangle
\label{D_half_q} \\
&= -\frac{i}{\mathcal{N}_{\Kn}^{(m)}}
\lim_{\qn \to 0} \pmb{\nabla}_q \langle \Psi_0 |\overline{Q}_{-\Kn-\frac{\qn}{2}} \overline{Q}_{\qn} \overline{Q}_{\Kn-\frac{\qn}{2}} | \Psi_0 \rangle.
\label{deriv_num}
\end{align}
In the second line of this, the normalizing factor has not been acted upon by $\pmb{\nabla}_{q}$ because it is even $\qn$, so that we may take $\qn \to 0$ directly in this term.  This is one of the simplifications that follows from rewriting $\mathcal{D}(\Kn)$ in the form shown in Eq. \ref{D_half_q}.

\subsection{QGD of the Magnetoplasmon in the SMA}

To carry out the concrete calculation, we use the techniques discussed in Refs. \onlinecite{Girvin_1986,Girvin_1984}, to derive, in a straightforward if somewhat involved calculation,
\begin{align}
\overline{Q}_{-\Kn-\frac{\qn}{2}}
\overline{Q}_{\qn}
\overline{Q}_{\Kn-\frac{\qn}{2}}
 & = \,
\overline{Q_{-\Kn-\frac{\qn}{2}} Q_{\qn} Q_{\Kn-\frac{\qn}{2}}}
+{1 \over 2}\left( q^*K\ell^2 -qK^*\ell^2 \right) \overline{Q_{-\Kn}Q_{\Kn}} \nonumber \\
&-\left( 1-e^{-(K^*+{q^* \over 2}) (K-{q \over 2}) \ell^2/2} \right) \overline{Q_{-\qn}Q_{\qn}} \nonumber \\
&-\overline{Q}_0 \left({1 \over 2}q^*K\ell^2 -{1 \over 2}qK^*\ell^2 \right)\left( 1-e^{-|\Kn|^2\ell^2/2} \right) + \mathcal{O}(q^2).
\label{GMP_expansion}
\end{align}
On the right-hand side of the above expression, wavevectors are written in complex notation (e.g., $q \equiv q_x - i q_y$), and the quantity $\overline{Q}_0$ is the same as the particle number, $\nu N_c$.

The reason for writing the product of the three projected operators on the left-hand side of Eq. \ref{GMP_expansion} in the form shown on the right-hand side is that,
upon taking expectation values, one obtains an explicit expression in terms of ground state correlation functions.
It is not difficult to show that
$\langle\Psi_0 |\overline{Q_{-\Kn-\frac{\qn}{2}} Q_{\qn} Q_{\Kn-\frac{\qn}{2}}} | \Psi_0 \rangle$
is even in $\qn$ for a circularly symmetric state, so this term does not contribute to Eq. \ref{deriv_num}.  The remaining terms involve density-density correlation functions, which may be expressed in terms of pair correlation functions.  In particular,
\begin{align}
\langle \Psi_0 | \overline{Q_{-\pn}Q_{\pn}} | \Psi_0 \rangle &= \int d^2 R_1 \int d^2 R_2
\sum_{i,j} \langle \Psi_0 | \delta(\Rn_1-\rn_i) \delta(\Rn_2-\rn_j) | \Psi_0 \rangle
e^{i\pn \cdot (\Rn_1 - \Rn_2)} \nonumber \\
&= \sum_i \int d^2 R_1 \int d^2 R_2 \Biggl\{
\sum_{i} \langle \Psi_0 | \delta(\Rn_1-\rn_i) | \Psi_0 \rangle \delta(\Rn_1-\Rn_2) \nonumber \\
&\quad + \sum_{i \ne j} \langle \Psi_0 | \delta(\Rn_1-\rn_i) \delta(\Rn_2-\rn_j) | \Psi_0 \rangle  \Biggr\} e^{i\pn \cdot (\Rn_1 - \Rn_2)} \nonumber \\
&= \nu N_c + n_0^2 \int d^2 R_1 \int d^2 R_2 e^{i\pn \cdot (\Rn_1 - \Rn_2)} g(\Rn_1-\Rn_2) \nonumber \\
&= \nu N_c + (2\pi)^2\nu N_c n_0 \delta(\pn) + \nu N_c n_0 \tilde{h}(\pn)
%
%
%\int d^2R e^{i\pn \cdot \Rn} h(\Rn)
\label{pair1}
\end{align}
where $n_0=2\pi\nu \ell^2$ is the particle density, $g(\Rn)$ is the ground state pair correlation function, and $h(\pn)=\int d^2R \left[g(\Rn)-1 \right] e^{i\pn \cdot \Rn}$, which has the property $\tilde{h}(0)=-1/n_0$.
%$\int d^2 R h(\Rn) = -1/n_0$.
Note in writing $g$ in terms of the difference in particle positions $\Rn_1 - \Rn_2$, we have assumed the infinite size limit has been taken, so that there is no impact from the system edge and the correlations are translationally invariant.

To compute the QGD we need to consider the expectation value $\langle \Psi_0 | \overline{Q}_{-\Kn-\frac{\qn}{2}} \overline{Q}_{\qn} \overline{Q}_{\Kn-\frac{\qn}{2}} | \Psi_0 \rangle$.  Some care must be taken in considering the contribution of $\langle \Psi_0 | \overline{Q_{-\qn}Q_{\qn}} | \Psi_0 \rangle$ in Eq. \ref{GMP_expansion}.  For small but finite $q$, Eq. \ref{pair1} shows this vanishes as $q^2$, so that this term does not contribute to the QGD.  The absence of any contribution from the $\delta(\qn)$ term comes about because the thermodynamic limit must be taken {\it before} taking $\qn \to 0$, which is required when we use the disk geometry to compute quantities that are translationally invariant in a system without an edge.

With these considerations, we arrive at
\begin{align}
\lim_{\qn \to 0}  \pmb{\nabla}_q \langle \Psi_0 |\overline{Q}_{-\Kn-\frac{\qn}{2}} \overline{Q}_{\qn} \overline{Q}_{\Kn-\frac{\qn}{2}} | \Psi_0 \rangle & = i\left(\Kn \times \hat{z} \ell^2 \right)
\left\{ \langle \Psi_0 | \overline{Q_{-\Kn} Q_{\Kn}} |\Psi_0  \rangle - \nu N_c \left( 1-e^{-|\Kn|^2\ell^2/2} \right) \right\} \nonumber \\
& = i\left(\Kn \times \hat{z} \ell^2 \right)
\left\{\nu N_c\left(1+ n_0 \tilde{h}(\Kn) \right)  - \nu N_c \left( 1-e^{-|\Kn|^2\ell^2/2} \right) \right\}.
\end{align}
The normalization $\mathcal{N}^{(m)}_{\Kn}$ for the excited state is calculated with similar methods \cite{Girvin_1986},
$$
\mathcal{N}^{(m)}_{\Kn}=
\langle \Psi_0 | \overline{Q}_{-\Kn} \overline{Q}_{\Kn} |\Psi_0 \rangle =
\langle \Psi_0 | \overline{Q_{-\Kn} Q}_{\Kn} |\Psi_0 \rangle + \nu N_{c}(e^{-|\Kn|^2\ell^2/2}-1),
$$
leading to the final result
\begin{equation}
\pmb{\mathcal{D}}(\Kn) = \left(\Kn \times \hat{z} \ell^2 \right)\frac{\nu N_c\left(1+ n_0 \tilde{h}(\Kn) \right)  + \nu N_c \left( e^{-|\Kn|^2\ell^2/2} - 1 \right) }{\nu N_c \left(1 + n_0 \tilde{h}(\Kn) + (e^{-|\Kn|^2\ell^2/2}-1) \right)}
= \Kn \times \hat{z} \ell^2.
\label{D_FQHE}
\end{equation}

\end{widetext}

We conclude this section with a few observations.  Firstly, we have arrived
at precisely the same relatively simple result that we found for the inter-Landau level exciton.  The more complicated internal correlations that play a role in intermediate steps of the analysis do not impact the final result.  We show below this is a result of the translationally invariance of the system.  Secondly, there is a quite different way to formulate the particle- and hole-hosting wavefunction spaces than the one we used.  This involves composite fermions \cite{Jain_book}.  In the mean-field approximation, the number of states in a composite fermion Landau level is equal to the number of particles, yielding a natural division between particle- and hole- hosting states.  However, to avoid introducing further approximations, it is necessary that the hole-hosting states be written in such a way that they all lie in the LLL of the original electron degrees of freedom.  Ultimately one arrives at Eq. \ref{final_D}, but the intermediate steps are more involved than what is presented here \cite{HAF_LB_unpub}.  Alternatively, one can adopt a further approximation in which the neutral excited state is simply a single particle-hole pair across two composite fermion Landau levels, with no LLL projection carried out.  This turns out to produce \cite{HAF_LB_unpub} a QGD the is $m$ times larger than that of Eq. \ref{D_FQHE}.  This reflects the large composite fermion magnetic length, and shows that sacrificing the projection of the state into the electron LLL introduces considerable error.

We now turn to this issue, to demonstrate that the simple result found in this section is an outcome of having a translationally invariant system, in a state that has a sharp momentum quantum number, and lies fully in the LLL.

\section{Dipole Moment in a Single Landau Level}

The result found for the magnetoplasmon above a Laughlin state is actually generic for any many-body state of a translationally-invariant system with well-defined momentum in a single Landau level.  In particular there is a relation between the center-of-mass (CM) position and the momentum.  To see this, we start with a many-body Hamiltonian in first quantization,
$$
H = \sum_j\frac{1}{2m}\left(\pn_j+\frac{e}{c} \An(\rn_j) \right)^2 + \sum_{i < j} v\left( \rn_j-\rn_j \right).
$$
Here it is convenient to work in Landau gauge, $\An(\rn)=B(0,x,0)$, $v(\rn)$  is an inter-electron potential, and $\pn_j$ is the momentum operator vector of the $j$th particle. In this gauge, states with a total momentum $K$ in the $\hat{y}$ direction are eigenstates of $\sum_j p_{j,y}$.  Consider the Landau level lowering operator for the CM degree of freedom,
$$
a_{CM} =  \frac{1}{\sqrt{2}} \sum_j\left[ \ell p_{j,x} - \frac{i}{\ell} \left( x_j + \ell^2 p_{j,y} \right) \right],
$$
which (by design) satisfies
$$
\left[H,a_{CM}\right] = -\frac{1}{m\ell^2} a_{CM}.
$$
If a state $\Psi_{K}$ with momentum $K\hat{y}$ resides fully in a single Landau level, we must have $\langle \Psi_{K} | a_{CM} | \Psi_K \rangle=\langle \Psi_{K} | a_{CM}^{\dag} | \Psi_K \rangle = 0$.  It follows
\begin{align}
\langle \Psi_K | \sum_j p_{j,x} | \Psi_K \rangle &\equiv \langle \Psi_K | P_x | \Psi_K \rangle = 0, \label{YPx} \\
\langle \Psi_K | \sum_j p_{j,y} + x_j/\ell^2 | \Psi_K \rangle &\equiv \langle \Psi_K | P_y + \hat{X}_{CM}/\ell^2 | \Psi_K \rangle = 0.
\label{XPy}
\end{align}
This gives the result
\begin{equation}
\langle \Psi_K | \hat{X}_{CM} | \Psi_K \rangle = -\ell^2 K,
\label{PX}
\end{equation}
which is one component of the result we obtained in the previous section.  For the other component, there is a subtlety.  In this formulation we should think of the system as having periodic boundary conditions in the $\hat{y}$ direction, and having edges in the $\hat{x}$ direction.  The state can in principle be compact in the latter dimension, so that the boundaries are very far away from the locations of the electrons.  The problem is that $\langle \Psi_K | \hat{Y}_{CM} | \Psi_K \rangle$ is not uniquely defined.

The electron density is uniform along the periodic direction of the cylinder, $\hat{y}$, which we take to have size $L_y$.  To uniquely define the $y_j$ coordinate, we fix an origin for coordinates of the positions $(x_j,y_j)$, and also a line at fixed ${x}_j$ for which $y_j$ jumps by $\pm L_y$ when particle $j$ passes through it.  The value of $\langle \Psi_K | \hat{Y}_{CM} | \Psi_K \rangle$ depends on the relative positions of the origin and this cut line.  However, whatever values we assign them, $\langle \Psi_K | Y_{CM} | \Psi_K \rangle$ will be the same for any $K$, because $\sum_j y_j$ cannot be localized anywhere on the cylinder if $\Psi_K$ is an eigenstate of $\sum_j p_{y,j}$.  We can remove the arbitrariness by choosing some reference state and considering only differences. Then
$$
\langle \Psi_K | \sum_j \rn_j | \Psi_K \rangle - \langle \Psi_0 | \sum_j \rn_j | \Psi_0 \rangle = -\hat{x} K\ell^2 = -\Kn \times \hat{z} \ell^2 .
$$
The overall $-$ sign is present because the shift is in the position of the electrons, which are negatively charged.
{Note that in our general formulation of the QGD, we also found it to be the {\it deviation} of the dipole moment from that of a reference state, $\Rn_0^{(K)}$.  The latter is not expected to have any $\Kn$ dependence in the thermodynamic limit for low-energy states, and, with a judicious choice of origin, can be made to vanish.}

We see then that the dipole moment for a lowest Landau level state with a momentum quantum number, for a system that is translationally invariant, {\it generically} takes the simple form found in the last section.  By contrast, we expect that had the translational invariance been broken, for example by a periodic potential, one would find deviations from this result.  This has been shown to be the case for particle-hole states in Landau levels of Dirac-like Hamiltonians \cite{Cao_PRB_2021}.

\section{Summary and Discussion}

Neutral excitations of a fermion system with a momentum quantum number carry an internal structure which is geometric in nature, a quantum geometric dipole, or QGD.  Previous work has demonstrated that this structure arises naturally for states that are well-described by particle-hole wavefunctions \cite{Cao_PRB_2021,Cao_PRL_2021,Cao_2022}.  In this work, we demonstrated that this concept is not tied to a particular form for the excited state wavefunction, and can be defined in a way that can be applied to states $|\Phi_{\Kn}\rangle$ of any form, provided they are labeled by a continuously varying wavevector $\Kn$.  The formulation exploits the density matrix associated with $|\Phi_{\Kn}\rangle$, allowing a set of single particle states to be defined.  These states are then divided into two groups, one of which is ``particle-hosting'', and the other ``hole-hosting,'' with the number of states in the latter
group equal to the number of fermions in the ground state.  These collections of states are then exploited to define quantities akin to Berry connections, whose difference is gauge invariant \cite{Cao_PRB_2021}, and is essentially the electric dipole moment of the excitation, up to a factor of the charge carried by the fermions.

To demonstrate that this formulation produces sensible results, we considered two concrete examples, both involving excitations of a two-dimensional electron gas in the quantum Hall regime.  In the first we considered a magnetoexciton above an integrally filled Landau level ground state.  We use the single mode approximation (SMA) to generate approximate excited state wavefunctions as a function of $\Kn$.  The result reproduces the QGD of an inter-Landau level exciton in the strong magnetic field limit, in which only two Landau levels (one for the hole, one for the electron) are involved in the state.  When substituted into semiclassical equations of motion \cite{Cao_PRB_2021},  this form is consistent with the Lorentz invariance that leads to drift motion of the exciton in the presence of an in-plane electric field.

In our second example, we considered magnetoplasmon excitations above a Laughlin state of a partially filled Landau level, again using the SMA.  In this the details of the calculation were more involved, in particular requiring a careful formulation of the division of states between particle- and hole-hosting.  This was accomplished by labeling single-particle states by wavevectors, allowing a definition of gradients with respect to $\Kn$ such that the particle-like and hole-like connections, $\pmb{\mathcal{A}}^{(p)}$ and $\pmb{\mathcal{A}}^{(h)}$, are well-defined.  The resulting formal expression for the QGD can be explicitly evaluated in terms of correlation functions in the ground state, but the final result does not involve these: one obtains precisely the result found for the magnetoexciton described above.  We demonstrated this simple result must emerge due to the combination of continuous translational invariance of the system, so that the state itself can be assigned a momentum along some direction, and the fact that the state lies fully in the lowest Landau level.

It is important to emphasize that the second example involves a state that cannot be written as a linear combination of single particle-hole pair states, demonstrating that to have a QGD, a state need not have this particular form.  And while we have applied our formulation to two examples of quantum Hall systems, this formulation is considerably more general than this, and can in principle be applied to any collection of neutral excitations with well-defined momenta, above some many-body ground state, in general dimensionality.  The challenge is that one needs explicit wavefunctions to carry through calculations.  It will be interesting to find further examples of states falling outside the single particle-hole paradigm for which the QGD may be computed.  Short of this, one can apply our formulation to approximate wavefunctions that involve small numbers of particle-hole pairs, for example as corrections to a single particle-hole wavefunction, to examine their impact on the dipole moment of the excitation.  Beyond computations of the QGD for different collective modes, it will be useful to find their equations of motion in applied electric and/or magnetic fields and determine how the QGD enters them \cite{Cao_PRB_2021}.  Indeed, the result above for the QGD of the magnetoplasmon above a Laughlin state suggests its equations of motion may be similar to those of a magnetoexciton above the integrally filled Landau level, in which case the former result can be again be understood as a consequence of Lorentz invariance in the underlying Hamiltonian, as is the case for the latter.
We leave these investigations for future work.

{\it Acknowledgements} -- LB was supported by Grant PID2021-125343NB-I00 (MCIN/AEI/FEDER, EU).  HAF acknowledges the support of the NSF through Grant
No. DMR-1914451, and thanks the Aspen Center for Physics (NSF
Grant No. 1066293) for its hospitality

\section*{Appendix A: {\bf K}-Dependence of ${\bf R}_0^{({\bf K})}$}

To understand how ${\bf R}_0^{({\bf K})}$ varies with ${\bf K}$ (cf. Eq. \ref{R0_def}), we examine its behavior for some excitations where we can generate approximate explicit wavefunctions.  As a first example, we consider plasmon excitations in a metal with a single band and a rotationally symmetric Fermi surface.  Within the RPA approximation, the wavefunction of such a state has the form \cite{Sawada:1957aa,Cao_PRL_2021}
\begin{equation}
|\Phi_{\bf K}\rangle = Q^{\dag} _{\Kn} |\Phi_{0}\rangle
\label{eqn:generic_excitation}
\end{equation}
where $|\Phi_0\rangle = \Pi_{q \le k_F} c_{\qn}^{\dag} |0\rangle$ is the approximate ground state, $k_F$ the Fermi wavevector, $|0\rangle$ the vacuum state, and
\begin{equation}
Q^{\dag} _{\Kn}=\sum _{|\qn| < k_F} \alpha_{\qn}(\Kn) c^{\dag} _{\qn +{\Kn}} c _{\qn}.
\label{wf0}
\end{equation}
The precise forms for the coefficients $\alpha_{\qn}(\Kn)$ may be written down, but are unimportant for our present purpose.  The ${\bf K}$-dependent density matrix in this case is
\begin{align}
\rho_{\Kn}({\bf r},{\bf r}') &= \langle \Phi_{n,{\bf K}}|\psi^{\dag}({\bf r}) \psi({\bf r}') |\Phi_{n,\bf {K}}\rangle \nonumber \\
%&= \sum_{{\bf q}_1,{\bf q}_2}
%\phi_{\qn_1}^{*}({\bf r})
%\phi_{\qn_2}({\bf r}')
%\langle \Phi_{n,{\bf K}}|c_{{\bf q}_1}^{\dag} c_{{\bf q}_2} |\Phi_{n,\bf {K}}\rangle \nonumber \\
%&= \sum_{{\bf q}_1} \phi^*_{\qn_1}({\bf r})\phi_{\qn_1}({\bf r}')
%\langle \Phi_{n,{\bf K}}|c_{{\bf q}_1}^{\dag} c_{{\bf q}_1} |\Phi_{n,\bf {K}}\rangle \nonumber \\
&\equiv \sum_{{\bf q}_1} \phi_{\qn_1}({\bf r})\phi^*_{\qn_1}({\bf r}') \lambda^{(\Kn)}_{\qn_1}. \nonumber
\end{align}
Here $\phi_{\qn}({\bf r})$ is the single particle state created by the operator $c_{{\bf q}}^{\dag}.$
We see that the eigenstates of $\rho_{\Kn}({\bf r},{\bf r}')$ are identical to those of $\rho_{\Kn=0}({\bf r},{\bf r}')$; all that changes is the average occupation of the single-particle states.  Moreover, it is not difficult to show that $\alpha_{\qn}(\Kn) \sim 1/\sqrt{N_c}$, where $N_c$ is the number of unit cells in the system, so that $\lambda^{(\Kn)}-\lambda^{(\Kn=0)} \sim 1/N_c$. Thus the change in the density matrix with $\Kn$ is negligibly small in the thermodynamic limit.  Independent of this last fact, the natural choice of hole-hosting states here is $\{ \phi_{\qn} \}$ with $q<k_F$, independent of $\Kn$.  It immediately follows that ${\bf R}_0^{({\bf K})}$ is independent of $\Kn$ for this system.  Moreover, it vanishes if the system is inversion symmetric.

As a second example we consider an insulating system whose ground state may be approximated by a single filled band, with quasiparticles which may reside in many higher energy bands, separated from the occupied band by a gap.  Within a Hartree-Fock approximation, each band $j$ hosts a set of single-particle states $\{ \phi_{j,\qn} \}$, and we assume the $j=0$ band is completely filled in the Hartree-Fock ground state $|\Phi_0\rangle$.
Low-energy neutral collective modes of such systems can typically be written in terms of some linear combination of particle-hole pairs, in the same form as Eq. \ref{eqn:generic_excitation}, but with
\begin{equation}
Q^{\dag} _{\Kn}=\sum _{\qn} \sum_{j>0} \alpha_{j,\qn}(\Kn) c^{\dag} _{j,\qn +{\Kn}} c_{0,\qn}.
\label{wf1}
\end{equation}
The coefficients $\alpha_{j,\qn}(\Kn)$ may be found within the RPA approximation, or within a time-dependent Hartree-Fock approximation.  In either case the precise form is again unimportant for our present purpose, except to note that $\alpha_{j,\qn}(\Kn) \sim 1/\sqrt{N_c}$.  In this case we have
\begin{align}
\rho_{\Kn}({\bf r},{\bf r}') &= \langle \Phi_{{\bf K}}|\psi^{\dag}({\bf r}) \psi({\bf r}') |\Phi_{\bf {K}}\rangle \nonumber \\
&= \sum_{{\bf q}} \sum_{j_1,j_2}\phi_{j_1,\qn}^*({\bf r})\phi_{j_2,\qn}({\bf r}')
\langle \Phi_{{\bf K}}|c_{j_1,{\bf q}}^{\dag} c_{j_2,{\bf q}} |\Phi_{\bf {K}}\rangle. \label{eqn:DMex2} \nonumber \\
%&\equiv \sum_{{\bf q}} \sum_{j_1,j_2} \phi_{j_1,\qn}^*({\bf r})\phi_{j_2,\qn}({\bf r}') \left[\lambda^{(\Kn)}_{0,\qn}\delta_{j_1,0}\delta_{j_2,0} + \alpha^*_{j_1,\qn}(\Kn)\alpha_{j_1,\qn}(\Kn)\right]. \nonumber
\end{align}
We need to compute
\begin{widetext}
\begin{equation}
\langle \Phi_{{\bf K}}|c_{j_1,{\bf q}}^{\dag} c_{j_2,{\bf q}} |\Phi_{\bf {K}}\rangle
=\sum_{\qn_1,\qn_2}\sum_{j_3,j_4 >0}\alpha^*_{j_3,\qn_1}(\Kn)\alpha_{j_4,\qn_2}(\Kn)
\langle \Phi_0 | c_{0,\qn_1}^{\dag} c_{j_3,\qn_1+\Kn} c^{\dag}_{\qn,j_1} c_{\qn,j_2} c^{\dag}_{j_4,\qn_2+\Kn} c_{0,\qn_2} |\Phi_0 \rangle.
\end{equation}
Taking into account the occupations of the bands, one finds
%Rewriting the matrix element in terms of normal-ordered operators gives
%\begin{align}
%\langle \Phi_0 | c_{0,\qn_1}^{\dag} c_{j_3,\qn_1+\Kn} c^{\dag}_{j_1,\qn} c_{j_2,\qn} c^{\dag}_{j_4,\qn_2+\Kn} c_{0,\qn_2} |\Phi_0 \rangle
%=&-\langle \Phi_0 | c_{0,\qn_1}^{\dag} c^{\dag}_{j_1,\qn} c^{\dag}_{j_4,\qn_2+\Kn} c_{j_3,\qn_1+\Kn} c_{j_2,\qn}  c_{0,\qn_2} |\Phi_0 \rangle \nonumber \\
%&+\langle \Phi_0 |c_{0,\qn_1}^{\dag} c^{\dag}_{j_1,\qn}  c_{j_2,\qn}  c_{0,\qn_2} |\Phi_0 \rangle
%\delta_{j_3,j_4}\delta_{\qn_1,\qn_2} \nonumber \\
%&-\langle \Phi_0 | c_{0,\qn_1}^{\dag} c^{\dag}_{j_1,\qn} c_{j_3,\qn_1+\Kn} c_{0,\qn_2} |\Phi_0 \rangle
%\delta_{j_2,j_4} \delta_{\qn,\qn_2+\Kn} \nonumber \\
%&-\langle \Phi_0 | c_{0,\qn_1}^{\dag} c^{\dag}_{j_4,\qn_2+\Kn} c_{j_2,\qn} c_{0,\qn_2} |\Phi_0 \rangle
%\delta_{j_1,j_3}\delta_{\qn,\qn_1+\Kn}\nonumber \\
%&+\langle \Phi_0 | c_{0,\qn_1}^{\dag} c_{0,\qn_2} |\Phi_0 \rangle
%\delta_{j_1,j_3}\delta_{j_2,j_4} \delta_{\qn,\qn_1+\Kn} \delta_{\qn,\qn_2+\Kn}
%\label{eqn:normal_order}
%\end{align}
%Because $|\Phi_0\rangle$ contains only electrons in the $j=0$ band, and $j_3,j_4>0$ in Eq. \ref{eqn:normal_order}, the first, third and fourth lines on the right-hand side of this equation vanish.  We are left with
\begin{align}
\langle \Phi_0 | c_{0,\qn_1}^{\dag} c_{j_3,\qn_1+\Kn} c^{\dag}_{j_1,\qn} c_{j_2,\qn} c^{\dag}_{j_4,\qn_2+\Kn} c_{0,\qn_2} |\Phi_0 \rangle
=
&\left[\delta_{j_1,0}\delta_{j_2,0}\delta_{\qn_1,\qn_2} -\delta_{j_1,0} \delta_{j_2,0} \delta_{\qn,\qn_1}\delta_{\qn,\qn_2}\right] \delta_{j_3,j_4}  \nonumber\\
&+ \delta_{\qn_1,\qn_2} \delta_{j_1,j_3}\delta_{j_2,j_4} \delta_{\qn,\qn_1+\Kn} \delta_{\qn,\qn_2+\Kn},
\end{align}
so that
\begin{equation}
\langle \Phi_{{\bf K}}|c_{j_1,{\bf q}}^{\dag} c_{j_2,{\bf q}} |\Phi_{\bf {K}}\rangle
=
\left[\sum_{j_3} \sum_{\qn_1 \ne \qn}|\alpha_{j_3,\qn_1}(\Kn)|^2 \right]\delta_{j_1,0}\delta_{j_2,0}
+ \alpha^*_{j_1,\qn-\Kn} \alpha_{j_2,\qn-\Kn}.
\label{eqn:ME_final}
\end{equation}
\end{widetext}
Eqs. \ref{eqn:DMex2} and \ref{eqn:ME_final} show that the $\Kn=0$ wavefunctions once again diagonalize the density matrix, up to corrections that vanish in the thermodynamic limit.

Eq. \ref{eqn:ME_final} points to what is needed for the density matrix to develop $\Kn$-dependence in the thermodynamic limit: one needs coherences between different bands that actually change with $\Kn$ within the excited state band.  Such excitations in general require rather high energy, because this implies correlations among the electrons that vary strongly away from those of the ground state.  In an excitation where the number of particle-hole pairs above the ground state needed to describe the state does not scale as the size of the system, the form of the hole-hosting states does not in practice change with $\Kn$ in the thermodynamic limit.

\begin{widetext}

\section*{Appendix B: Calculation of $\sum_{n>0} \Gamma_n(\Kn_1,\Kn_2)$ and $\Gamma_0(\Kn_1,\Kn_2)$}

In this Appendix, we present some details of the calculations involved in the computation of the QGD for a magnetoexciton state above a single filled Landau level, with the state generated by the single-mode approximation (SMA).  In Section \ref{sec:filledLL}, one finds an expression for the QGD, Eq. \ref{eqn:D}, that involves the quantities $\Gamma_n$ defined in Eqs. \ref{eqn:Gamma_n} and \ref{eqn:Gamma_0}.  These latter quantities involve matrix elements that can be explicitly evaluated.  Specifically, one requires
\begin{equation}
\langle \Phi_{\Kn_1} | c^{\dag}_{n,X_1} c_{n,X_2} | \Phi_{\Kn_2} \rangle
= \frac{1}{\sqrt{\mathcal{N}_{\Kn_1}\mathcal{N}_{\Kn_2}}} \langle \Phi_0 | Q_{\Kn_1} c^{\dag}_{n,X_1} c_{n,X_2} Q_{\Kn_2}^{\dag} | \Phi_0 \rangle,
\nonumber
\end{equation}
with $Q_{\Kn}^{\dag} = \sum_{n_1,n_2} \sum_{X_1,X_2} \langle \phi_{n_1,X_1} | e^{i\Kn \cdot \rn} | \phi_{n_2,X_2} \rangle c^{\dag}_{n_1,X_1} c_{n_2,X_2}$.
Noting that, for $n>0$,
\begin{align}
c_{n,X} Q_{\Kn}^{\dag} | \Phi_0 \rangle &= \sum_{X_1,X_2} \sum_{n_1,n_2} \langle \phi_{n_1,X_1} | e^{i\Kn \cdot \rn} | \phi_{n_2,X_2} \rangle c_{n,X} c^{\dag}_{n_1,X_1} c_{n_2,X_2} | \Phi_0 \rangle \nonumber \\
&= \sum_{X_1,X_2} \sum_{n_1} \langle \phi_{n_1,X_1} | e^{i\Kn \cdot \rn} | \phi_{0,X_2} \rangle c_{n,X} c^{\dag}_{n_1,X_1} c_{0,X_2} | \Phi_0 \rangle \nonumber \\
&= \sum_{X_2} \langle \phi_{n,X} | e^{i\Kn \cdot \rn} | \phi_{0,X_2} \rangle c_{0,X_2} | \Phi_0 \rangle, \nonumber
\end{align}
we arrive at
\begin{equation}
\langle \Phi_0 | Q_{\Kn_1} c^{\dag}_{n,X_1} c_{n,X_2} Q_{\Kn_2}^{\dag} | \Phi_0 \rangle =
\sum_{X_3} \langle \phi_{n,X_2} | e^{i \Kn_2 \cdot \rn} | \phi_{0,X_3} \rangle
\langle \phi_{0,X_3} | e^{-i \Kn_1 \cdot \rn} | \phi_{n,X_1} \rangle.
\label{eqn:ME1}
\end{equation}
A similar calculation yields
\begin{equation}
\langle \Phi_0 | Q_{\Kn_1} c_{0,X_1} c_{0,X_2}^{\dag} Q_{\Kn_2}^{\dag} | \Phi_0 \rangle =
\sum_{n_3 > 0}\sum_{X_3} \langle \phi_{0,X_1} | e^{-i \Kn_1 \cdot \rn} | \phi_{n_3,X_3} \rangle
\langle \phi_{n_3,X_3} | e^{i \Kn_2 \cdot \rn} | \phi_{0,X_2} \rangle.
\label{eqn:ME2}
\end{equation}

The quantities of interest can now be written as
\begin{align}
\sum_{n>0}\Gamma_n(\Kn_1,\Kn_2) =&
\left[ \frac{1}{\mathcal{N}_{\Kn_1} \mathcal{N}_{\Kn_2}} \right]^{1/2}
\sum_{X_1,X_2,X_3}\sum_{n > 0}
\langle \phi_{n,X_2} | e^{i\Kn_2 \cdot \rn} | \phi_{0,X_3} \rangle
\langle \phi_{0,X_3} | e^{-i\Kn_1 \cdot \rn} | \phi_{n,X_1} \rangle
\langle \phi_{n,X_1} | e^{i(\Kn_1-\Kn_2) \cdot \rn} | \phi_{n,X_2} \rangle, \nonumber \\
\Gamma_0(\Kn_1,\Kn_2) =&
\left[ \frac{1}{\mathcal{N}_{\Kn_1} \mathcal{N}_{\Kn_2}} \right]^{1/2}
\sum_{X_1,X_2,X_3}\sum_{n>0}
\langle \phi_{0,X_1} | e^{-i\Kn_1 \cdot \rn} | \phi_{n,X_3} \rangle
\langle \phi_{n,X_3} | e^{i\Kn_2 \cdot \rn} | \phi_{n,X_2} \rangle
\langle \phi_{0,X_2} | e^{i(\Kn_1-\Kn_2) \cdot \rn} | \phi_{0,X_1} \rangle.
\nonumber
\end{align}

This expression can be evaluated explicitly with use of the matrix element
\begin{equation}
\langle \phi_{n',X'} | e^{i\qn \cdot \rn} | \phi_{n,X} \rangle
=
e^{iq_x(X+X')} \delta_{X',X-q_y\ell^2}
\left[ \frac{n!}{n'!} \right]^{1/2}
\left[\frac{(q_y +i q_x)\ell}{\sqrt{2}} \right]^{n'-n}
e^{-q^2\ell^2/4} L_n^{n'-n}\left[ \frac{q^2\ell^2}{2} \right],
\nonumber
\end{equation}
where $L_n^{n'-n}$ is an associated Laguerre polynomial, and in writing this we have assumed $n' \ge n$.
With some algebra, it is possible to show
\begin{align}
\sum_{X_1,X_2,X_3} &
\langle \phi_{n,X_2} | e^{i\Kn_2 \cdot \rn} | \phi_{0,X_3} \rangle
\langle \phi_{0,X_3} | e^{-i\Kn_1 \cdot \rn} | \phi_{n,X_1} \rangle
\langle \phi_{n,X_1} | e^{i(\Kn_1-\Kn_2) \cdot \rn} | \phi_{n,X_2} \rangle \nonumber \\
&=
\frac{g}{n!} e^{-i\hat{z} \cdot (\Kn_1 \times \Kn_2) \ell^2}
\left[ \frac{(K_{2y} + iK_{2x})\ell}{\sqrt{2}} \right]^n
\left[ \frac{(K_{1y} - iK_{1x})\ell}{\sqrt{2}} \right]^n
L_n^0\left[\frac{(\Kn_1-\Kn_2)^2\ell^2}{2} \right] \nonumber \\
&\times e^{-K_1^2\ell^2/4-K_2^2\ell^2/4}e^{-(\Kn_1-\Kn_2)^2\ell^2/4}, \nonumber
\end{align}
where $g$ is the degeneracy of a Landau level.  Noting that $L_n^0(x) \approx 1-nx$ for small $x$, we can set $L_n^0\left[\frac{(\Kn_1-\Kn_2)^2\ell^2}{2} \right]\to 1$ without incurring any error in the final answer, since we take only a single gradient and then set $\Kn_2 \to \Kn_1$.  The last Gaussian factor may be dropped for the same reason.

With these results, we can write
\begin{align}
\sum_{n>0} \Gamma_n(\Kn_1,\Kn_2) = &\frac{g}{\sqrt{\mathcal{N}_{\Kn_1}\mathcal{N}_{\Kn_2}}} e^{-i\hat{z} \cdot (\Kn_1 \times \Kn_2)\ell^2/2 - K_1^2\ell^2/4 - K_2^2\ell^2/4} \nonumber \\
&\times \left\{ \exp \left[ \frac{(K_{2y}+iK_{2x})(K_{1y}-iK_{1x})\ell^2}{2} \right] -1 \right\} + \mathcal{O}(\delta K)^2 \nonumber \\
= &\frac{g}{\sqrt{\mathcal{N}_{\Kn_1}\mathcal{N}_{\Kn_2}}}\left\{e^{-i\hat{z} \cdot (\Kn_1 \times \Kn_2)\ell^2} - e^{-i\hat{z} \cdot (\Kn_1 \times \Kn_2)\ell^2/2 -K_1^2\ell^2/4 - K_2^2\ell^2/4} \right\}
+ \mathcal{O}(\delta K)^2,
\nonumber
\end{align}
where $\delta \Kn \equiv \Kn_1 - \Kn_2$.  A similar calculation yields the result
\begin{equation*}
\Gamma_0(\Kn_1,\Kn_2) = \frac{g}{\sqrt{\mathcal{N}_{\Kn_1}\mathcal{N}_{\Kn_2}}}
\left\{
1-e^{i\hat{z} \cdot (\Kn_1 \times \Kn_2)\ell^2/2 -K_1^2\ell^2/4 - K_2^2\ell^2/4}
\right\}
+ \mathcal{O}(\delta K)^2.
\end{equation*}

\end{widetext}


\begin{thebibliography}{66}%
\makeatletter
\providecommand \@ifxundefined [1]{%
 \@ifx{#1\undefined}
}%
\providecommand \@ifnum [1]{%
 \ifnum #1\expandafter \@firstoftwo
 \else \expandafter \@secondoftwo
 \fi
}%
\providecommand \@ifx [1]{%
 \ifx #1\expandafter \@firstoftwo
 \else \expandafter \@secondoftwo
 \fi
}%
\providecommand \natexlab [1]{#1}%
\providecommand \enquote  [1]{``#1''}%
\providecommand \bibnamefont  [1]{#1}%
\providecommand \bibfnamefont [1]{#1}%
\providecommand \citenamefont [1]{#1}%
\providecommand \href@noop [0]{\@secondoftwo}%
\providecommand \href [0]{\begingroup \@sanitize@url \@href}%
\providecommand \@href[1]{\@@startlink{#1}\@@href}%
\providecommand \@@href[1]{\endgroup#1\@@endlink}%
\providecommand \@sanitize@url [0]{\catcode `\\12\catcode `\$12\catcode
  `\&12\catcode `\#12\catcode `\^12\catcode `\_12\catcode `\%12\relax}%
\providecommand \@@startlink[1]{}%
\providecommand \@@endlink[0]{}%
\providecommand \url  [0]{\begingroup\@sanitize@url \@url }%
\providecommand \@url [1]{\endgroup\@href {#1}{\urlprefix }}%
\providecommand \urlprefix  [0]{URL }%
\providecommand \Eprint [0]{\href }%
\providecommand \doibase [0]{https://doi.org/}%
\providecommand \selectlanguage [0]{\@gobble}%
\providecommand \bibinfo  [0]{\@secondoftwo}%
\providecommand \bibfield  [0]{\@secondoftwo}%
\providecommand \translation [1]{[#1]}%
\providecommand \BibitemOpen [0]{}%
\providecommand \bibitemStop [0]{}%
\providecommand \bibitemNoStop [0]{.\EOS\space}%
\providecommand \EOS [0]{\spacefactor3000\relax}%
\providecommand \BibitemShut  [1]{\csname bibitem#1\endcsname}%
\let\auto@bib@innerbib\@empty
%</preamble>
\bibitem [{\citenamefont {Thouless}\ \emph {et~al.}(1982)\citenamefont
  {Thouless}, \citenamefont {Kohmoto}, \citenamefont {Nightingale},\ and\
  \citenamefont {den Nijs}}]{Thouless_1982}%
  \BibitemOpen
  \bibfield  {author} {\bibinfo {author} {\bibfnamefont {D.~J.}\ \bibnamefont
  {Thouless}}, \bibinfo {author} {\bibfnamefont {M.}~\bibnamefont {Kohmoto}},
  \bibinfo {author} {\bibfnamefont {M.~P.}\ \bibnamefont {Nightingale}},\ and\
  \bibinfo {author} {\bibfnamefont {M.}~\bibnamefont {den Nijs}},\ }\bibfield
  {title} {\bibinfo {title} {Quantized hall conductance in a two-dimensional
  periodic potential},\ }\href {https://doi.org/10.1103/PhysRevLett.49.405}
  {\bibfield  {journal} {\bibinfo  {journal} {Phys. Rev. Lett.}\ }\textbf
  {\bibinfo {volume} {49}},\ \bibinfo {pages} {405} (\bibinfo {year}
  {1982})}\BibitemShut {NoStop}%
\bibitem [{\citenamefont {Avron}\ \emph {et~al.}(2003)\citenamefont {Avron},
  \citenamefont {Osadchy},\ and\ \citenamefont {Seiler}}]{Avron_2003}%
  \BibitemOpen
  \bibfield  {author} {\bibinfo {author} {\bibfnamefont {J.~E.}\ \bibnamefont
  {Avron}}, \bibinfo {author} {\bibfnamefont {D.}~\bibnamefont {Osadchy}},\
  and\ \bibinfo {author} {\bibfnamefont {R.}~\bibnamefont {Seiler}},\
  }\bibfield  {title} {\bibinfo {title} {{A Topological Look at the Quantum
  Hall Effect}},\ }\href {https://doi.org/10.1063/1.1611351} {\bibfield
  {journal} {\bibinfo  {journal} {Physics Today}\ }\textbf {\bibinfo {volume}
  {56}},\ \bibinfo {pages} {38} (\bibinfo {year} {2003})},\ \Eprint
  {https://arxiv.org/abs/https://pubs.aip.org/physicstoday/article-pdf/56/8/38/16651528/38\_1\_online.pdf}
  {https://pubs.aip.org/physicstoday/article-pdf/56/8/38/16651528/38\_1\_online.pdf}
  \BibitemShut {NoStop}%
\bibitem [{\citenamefont {Girvin}\ and\ \citenamefont
  {Yang}(2019)}]{Girvin-Book}%
  \BibitemOpen
  \bibfield  {author} {\bibinfo {author} {\bibfnamefont {S.~M.}\ \bibnamefont
  {Girvin}}\ and\ \bibinfo {author} {\bibfnamefont {K.}~\bibnamefont {Yang}},\
  }\href@noop {} {\emph {\bibinfo {title} {Modern Condensed Matter Physics}}}\
  (\bibinfo  {publisher} {Cambridge University Press},\ \bibinfo {address}
  {Cambridge},\ \bibinfo {year} {2019})\BibitemShut {NoStop}%
\bibitem [{\citenamefont {Haldane}(1988)}]{Haldane_1988}%
  \BibitemOpen
  \bibfield  {author} {\bibinfo {author} {\bibfnamefont {F.~D.~M.}\
  \bibnamefont {Haldane}},\ }\bibfield  {title} {\bibinfo {title} {Model for a
  quantum hall effect without landau levels: Condensed-matter realization of
  the "parity anomaly"},\ }\href {https://doi.org/10.1103/PhysRevLett.61.2015}
  {\bibfield  {journal} {\bibinfo  {journal} {Phys. Rev. Lett.}\ }\textbf
  {\bibinfo {volume} {61}},\ \bibinfo {pages} {2015} (\bibinfo {year}
  {1988})}\BibitemShut {NoStop}%
\bibitem [{\citenamefont {Kane}\ and\ \citenamefont {Mele}(2005)}]{Kane_2005}%
  \BibitemOpen
  \bibfield  {author} {\bibinfo {author} {\bibfnamefont {C.~L.}\ \bibnamefont
  {Kane}}\ and\ \bibinfo {author} {\bibfnamefont {E.~J.}\ \bibnamefont
  {Mele}},\ }\bibfield  {title} {\bibinfo {title} {${Z}_{2}$ topological order
  and the quantum spin hall effect},\ }\href
  {https://doi.org/10.1103/PhysRevLett.95.146802} {\bibfield  {journal}
  {\bibinfo  {journal} {Phys. Rev. Lett.}\ }\textbf {\bibinfo {volume} {95}},\
  \bibinfo {pages} {146802} (\bibinfo {year} {2005})}\BibitemShut {NoStop}%
\bibitem [{\citenamefont {Bernevig}\ \emph {et~al.}(2006)\citenamefont
  {Bernevig}, \citenamefont {Hughes},\ and\ \citenamefont
  {Zhang}}]{Bernevig_2006}%
  \BibitemOpen
  \bibfield  {author} {\bibinfo {author} {\bibfnamefont {B.~A.}\ \bibnamefont
  {Bernevig}}, \bibinfo {author} {\bibfnamefont {T.~L.}\ \bibnamefont
  {Hughes}},\ and\ \bibinfo {author} {\bibfnamefont {S.-C.}\ \bibnamefont
  {Zhang}},\ }\bibfield  {title} {\bibinfo {title} {Quantum spin hall effect
  and topological phase transition in hgte quantum wells},\ }\href
  {https://doi.org/10.1126/science.1133734} {\bibfield  {journal} {\bibinfo
  {journal} {Science}\ }\textbf {\bibinfo {volume} {314}},\ \bibinfo {pages}
  {1757} (\bibinfo {year} {2006})},\ \Eprint
  {https://arxiv.org/abs/https://www.science.org/doi/pdf/10.1126/science.1133734}
  {https://www.science.org/doi/pdf/10.1126/science.1133734} \BibitemShut
  {NoStop}%
\bibitem [{\citenamefont {Qi}\ and\ \citenamefont {Zhang}(2010)}]{Qi_2010}%
  \BibitemOpen
  \bibfield  {author} {\bibinfo {author} {\bibfnamefont {X.-L.}\ \bibnamefont
  {Qi}}\ and\ \bibinfo {author} {\bibfnamefont {S.-C.}\ \bibnamefont {Zhang}},\
  }\bibfield  {title} {\bibinfo {title} {{The quantum spin Hall effect and
  topological insulators}},\ }\href {https://doi.org/10.1063/1.3293411}
  {\bibfield  {journal} {\bibinfo  {journal} {Physics Today}\ }\textbf
  {\bibinfo {volume} {63}},\ \bibinfo {pages} {33} (\bibinfo {year} {2010})},\
  \Eprint
  {https://arxiv.org/abs/https://pubs.aip.org/physicstoday/article-pdf/63/1/33/9879625/33\_1\_online.pdf}
  {https://pubs.aip.org/physicstoday/article-pdf/63/1/33/9879625/33\_1\_online.pdf}
  \BibitemShut {NoStop}%
\bibitem [{\citenamefont {Nagaosa}\ \emph {et~al.}(2010)\citenamefont
  {Nagaosa}, \citenamefont {Sinova}, \citenamefont {Onoda}, \citenamefont
  {MacDonald},\ and\ \citenamefont {Ong}}]{Nagaosa_2010}%
  \BibitemOpen
  \bibfield  {author} {\bibinfo {author} {\bibfnamefont {N.}~\bibnamefont
  {Nagaosa}}, \bibinfo {author} {\bibfnamefont {J.}~\bibnamefont {Sinova}},
  \bibinfo {author} {\bibfnamefont {S.}~\bibnamefont {Onoda}}, \bibinfo
  {author} {\bibfnamefont {A.~H.}\ \bibnamefont {MacDonald}},\ and\ \bibinfo
  {author} {\bibfnamefont {N.~P.}\ \bibnamefont {Ong}},\ }\bibfield  {title}
  {\bibinfo {title} {Anomalous hall effect},\ }\href
  {https://doi.org/10.1103/RevModPhys.82.1539} {\bibfield  {journal} {\bibinfo
  {journal} {Rev. Mod. Phys.}\ }\textbf {\bibinfo {volume} {82}},\ \bibinfo
  {pages} {1539} (\bibinfo {year} {2010})}\BibitemShut {NoStop}%
\bibitem [{\citenamefont {Sundaram}\ and\ \citenamefont
  {Niu}(1999)}]{Sundaram_1999}%
  \BibitemOpen
  \bibfield  {author} {\bibinfo {author} {\bibfnamefont {G.}~\bibnamefont
  {Sundaram}}\ and\ \bibinfo {author} {\bibfnamefont {Q.}~\bibnamefont {Niu}},\
  }\bibfield  {title} {\bibinfo {title} {Wave-packet dynamics in slowly
  perturbed crystals: Gradient corrections and berry-phase effects},\ }\href
  {https://doi.org/10.1103/PhysRevB.59.14915} {\bibfield  {journal} {\bibinfo
  {journal} {Phys. Rev. B}\ }\textbf {\bibinfo {volume} {59}},\ \bibinfo
  {pages} {14915} (\bibinfo {year} {1999})}\BibitemShut {NoStop}%
\bibitem [{\citenamefont {Haldane}(2004)}]{Haldane_2004}%
  \BibitemOpen
  \bibfield  {author} {\bibinfo {author} {\bibfnamefont {F.~D.~M.}\
  \bibnamefont {Haldane}},\ }\bibfield  {title} {\bibinfo {title} {Berry
  curvature on the fermi surface: Anomalous hall effect as a topological
  fermi-liquid property},\ }\href
  {https://doi.org/10.1103/PhysRevLett.93.206602} {\bibfield  {journal}
  {\bibinfo  {journal} {Phys. Rev. Lett.}\ }\textbf {\bibinfo {volume} {93}},\
  \bibinfo {pages} {206602} (\bibinfo {year} {2004})}\BibitemShut {NoStop}%
\bibitem [{\citenamefont {Xiao}\ \emph {et~al.}(2010)\citenamefont {Xiao},
  \citenamefont {Chang},\ and\ \citenamefont {Niu}}]{Xiao_2010}%
  \BibitemOpen
  \bibfield  {author} {\bibinfo {author} {\bibfnamefont {D.}~\bibnamefont
  {Xiao}}, \bibinfo {author} {\bibfnamefont {M.-C.}\ \bibnamefont {Chang}},\
  and\ \bibinfo {author} {\bibfnamefont {Q.}~\bibnamefont {Niu}},\ }\bibfield
  {title} {\bibinfo {title} {Berry phase effects on electronic properties},\
  }\href {https://doi.org/10.1103/RevModPhys.82.1959} {\bibfield  {journal}
  {\bibinfo  {journal} {Rev. Mod. Phys.}\ }\textbf {\bibinfo {volume} {82}},\
  \bibinfo {pages} {1959} (\bibinfo {year} {2010})}\BibitemShut {NoStop}%
\bibitem [{\citenamefont {Lapa}\ and\ \citenamefont
  {Hughes}(2019)}]{Lapa_2019}%
  \BibitemOpen
  \bibfield  {author} {\bibinfo {author} {\bibfnamefont {M.~F.}\ \bibnamefont
  {Lapa}}\ and\ \bibinfo {author} {\bibfnamefont {T.~L.}\ \bibnamefont
  {Hughes}},\ }\bibfield  {title} {\bibinfo {title} {Semiclassical wave packet
  dynamics in nonuniform electric fields},\ }\href
  {https://doi.org/10.1103/PhysRevB.99.121111} {\bibfield  {journal} {\bibinfo
  {journal} {Phys. Rev. B}\ }\textbf {\bibinfo {volume} {99}},\ \bibinfo
  {pages} {121111} (\bibinfo {year} {2019})}\BibitemShut {NoStop}%
\bibitem [{\citenamefont {T\"orm\"a}(2023)}]{Torma_2023}%
  \BibitemOpen
  \bibfield  {author} {\bibinfo {author} {\bibfnamefont {P.}~\bibnamefont
  {T\"orm\"a}},\ }\bibfield  {title} {\bibinfo {title} {Essay: Where can
  quantum geometry lead us?},\ }\href
  {https://doi.org/10.1103/PhysRevLett.131.240001} {\bibfield  {journal}
  {\bibinfo  {journal} {Phys. Rev. Lett.}\ }\textbf {\bibinfo {volume} {131}},\
  \bibinfo {pages} {240001} (\bibinfo {year} {2023})}\BibitemShut {NoStop}%
\bibitem [{\citenamefont {Sodemann}\ and\ \citenamefont
  {Fu}(2015)}]{Sodemann_2015}%
  \BibitemOpen
  \bibfield  {author} {\bibinfo {author} {\bibfnamefont {I.}~\bibnamefont
  {Sodemann}}\ and\ \bibinfo {author} {\bibfnamefont {L.}~\bibnamefont {Fu}},\
  }\bibfield  {title} {\bibinfo {title} {Quantum nonlinear hall effect induced
  by berry curvature dipole in time-reversal invariant materials},\ }\href
  {https://doi.org/10.1103/PhysRevLett.115.216806} {\bibfield  {journal}
  {\bibinfo  {journal} {Phys. Rev. Lett.}\ }\textbf {\bibinfo {volume} {115}},\
  \bibinfo {pages} {216806} (\bibinfo {year} {2015})}\BibitemShut {NoStop}%
\bibitem [{\citenamefont {Ortix}(2021)}]{Ortix_2021}%
  \BibitemOpen
  \bibfield  {author} {\bibinfo {author} {\bibfnamefont {C.}~\bibnamefont
  {Ortix}},\ }\bibfield  {title} {\bibinfo {title} {Nonlinear hall effect with
  time-reversal symmetry: Theory and material realizations},\ }\href@noop {}
  {\bibfield  {journal} {\bibinfo  {journal} {Advanced Quantum Technologies}\
  }\textbf {\bibinfo {volume} {4}},\ \bibinfo {pages} {2100056} (\bibinfo
  {year} {2021})}\BibitemShut {NoStop}%
\bibitem [{\citenamefont {Nagaosa}\ and\ \citenamefont
  {Yanase}(2024)}]{Nagaosa_2024}%
  \BibitemOpen
  \bibfield  {author} {\bibinfo {author} {\bibfnamefont {N.}~\bibnamefont
  {Nagaosa}}\ and\ \bibinfo {author} {\bibfnamefont {Y.}~\bibnamefont
  {Yanase}},\ }\bibfield  {title} {\bibinfo {title} {Nonreciprocal transport
  and optical phenomena in quantum materials},\ }\href
  {https://doi.org/https://doi.org/10.1146/annurev-conmatphys-032822-033734}
  {\bibfield  {journal} {\bibinfo  {journal} {Annual Review of Condensed Matter
  Physics}\ }\textbf {\bibinfo {volume} {15}},\ \bibinfo {pages} {63} (\bibinfo
  {year} {2024})}\BibitemShut {NoStop}%
\bibitem [{\citenamefont {Xu}(2024)}]{Xu_2024}%
  \BibitemOpen
  \bibfield  {author} {\bibinfo {author} {\bibfnamefont {S.-Y.}\ \bibnamefont
  {Xu}},\ }\bibfield  {title} {\bibinfo {title} {A new hall effect from quantum
  geometry},\ }\href@noop {} {\bibfield  {journal} {\bibinfo  {journal}
  {Physics}\ }\textbf {\bibinfo {volume} {17}},\ \bibinfo {pages} {38}
  (\bibinfo {year} {2024})}\BibitemShut {NoStop}%
\bibitem [{\citenamefont {Wang}\ \emph {et~al.}(2024)\citenamefont {Wang},
  \citenamefont {Zhu}, \citenamefont {Chen}, \citenamefont {Wang},
  \citenamefont {Liu}, \citenamefont {Huang}, \citenamefont {Jiang},
  \citenamefont {Zhao}, \citenamefont {Shi}, \citenamefont {Tian},
  \citenamefont {Wang}, \citenamefont {Yao}, \citenamefont {Yu}, \citenamefont
  {Wang}, \citenamefont {Xiao}, \citenamefont {Yang},\ and\ \citenamefont
  {Wu}}]{Wang_2024}%
  \BibitemOpen
  \bibfield  {author} {\bibinfo {author} {\bibfnamefont {L.}~\bibnamefont
  {Wang}}, \bibinfo {author} {\bibfnamefont {J.}~\bibnamefont {Zhu}}, \bibinfo
  {author} {\bibfnamefont {H.}~\bibnamefont {Chen}}, \bibinfo {author}
  {\bibfnamefont {H.}~\bibnamefont {Wang}}, \bibinfo {author} {\bibfnamefont
  {J.}~\bibnamefont {Liu}}, \bibinfo {author} {\bibfnamefont {Y.-X.}\
  \bibnamefont {Huang}}, \bibinfo {author} {\bibfnamefont {B.}~\bibnamefont
  {Jiang}}, \bibinfo {author} {\bibfnamefont {J.}~\bibnamefont {Zhao}},
  \bibinfo {author} {\bibfnamefont {H.}~\bibnamefont {Shi}}, \bibinfo {author}
  {\bibfnamefont {G.}~\bibnamefont {Tian}}, \bibinfo {author} {\bibfnamefont
  {H.}~\bibnamefont {Wang}}, \bibinfo {author} {\bibfnamefont {Y.}~\bibnamefont
  {Yao}}, \bibinfo {author} {\bibfnamefont {D.}~\bibnamefont {Yu}}, \bibinfo
  {author} {\bibfnamefont {Z.}~\bibnamefont {Wang}}, \bibinfo {author}
  {\bibfnamefont {C.}~\bibnamefont {Xiao}}, \bibinfo {author} {\bibfnamefont
  {S.~A.}\ \bibnamefont {Yang}},\ and\ \bibinfo {author} {\bibfnamefont
  {X.}~\bibnamefont {Wu}},\ }\bibfield  {title} {\bibinfo {title} {Orbital
  magneto-nonlinear anomalous hall effect in kagome magnet
  ${\mathrm{fe}}_{3}{\mathrm{sn}}_{2}$},\ }\href
  {https://doi.org/10.1103/PhysRevLett.132.106601} {\bibfield  {journal}
  {\bibinfo  {journal} {Phys. Rev. Lett.}\ }\textbf {\bibinfo {volume} {132}},\
  \bibinfo {pages} {106601} (\bibinfo {year} {2024})}\BibitemShut {NoStop}%
\bibitem [{\citenamefont {Peotta}\ and\ \citenamefont
  {T{\"o}rm{\"a}}(2015)}]{Peotta_2015}%
  \BibitemOpen
  \bibfield  {author} {\bibinfo {author} {\bibfnamefont {S.}~\bibnamefont
  {Peotta}}\ and\ \bibinfo {author} {\bibfnamefont {P.}~\bibnamefont
  {T{\"o}rm{\"a}}},\ }\bibfield  {title} {\bibinfo {title} {Superfluidity in
  topologically nontrivial flat bands},\ }\href
  {https://doi.org/10.1038/ncomms9944} {\bibfield  {journal} {\bibinfo
  {journal} {Nature Communications}\ }\textbf {\bibinfo {volume} {6}},\
  \bibinfo {pages} {8944} (\bibinfo {year} {2015})}\BibitemShut {NoStop}%
\bibitem [{\citenamefont {Julku}\ \emph {et~al.}(2016)\citenamefont {Julku},
  \citenamefont {Peotta}, \citenamefont {Vanhala}, \citenamefont {Kim},\ and\
  \citenamefont {T\"orm\"a}}]{Julku_2016}%
  \BibitemOpen
  \bibfield  {author} {\bibinfo {author} {\bibfnamefont {A.}~\bibnamefont
  {Julku}}, \bibinfo {author} {\bibfnamefont {S.}~\bibnamefont {Peotta}},
  \bibinfo {author} {\bibfnamefont {T.~I.}\ \bibnamefont {Vanhala}}, \bibinfo
  {author} {\bibfnamefont {D.-H.}\ \bibnamefont {Kim}},\ and\ \bibinfo {author}
  {\bibfnamefont {P.}~\bibnamefont {T\"orm\"a}},\ }\bibfield  {title} {\bibinfo
  {title} {Geometric origin of superfluidity in the lieb-lattice flat band},\
  }\href {https://doi.org/10.1103/PhysRevLett.117.045303} {\bibfield  {journal}
  {\bibinfo  {journal} {Phys. Rev. Lett.}\ }\textbf {\bibinfo {volume} {117}},\
  \bibinfo {pages} {045303} (\bibinfo {year} {2016})}\BibitemShut {NoStop}%
\bibitem [{\citenamefont {Liang}\ \emph {et~al.}(2017)\citenamefont {Liang},
  \citenamefont {Vanhala}, \citenamefont {Peotta}, \citenamefont {Siro},
  \citenamefont {Harju},\ and\ \citenamefont {T\"orm\"a}}]{Liang_2017}%
  \BibitemOpen
  \bibfield  {author} {\bibinfo {author} {\bibfnamefont {L.}~\bibnamefont
  {Liang}}, \bibinfo {author} {\bibfnamefont {T.~I.}\ \bibnamefont {Vanhala}},
  \bibinfo {author} {\bibfnamefont {S.}~\bibnamefont {Peotta}}, \bibinfo
  {author} {\bibfnamefont {T.}~\bibnamefont {Siro}}, \bibinfo {author}
  {\bibfnamefont {A.}~\bibnamefont {Harju}},\ and\ \bibinfo {author}
  {\bibfnamefont {P.}~\bibnamefont {T\"orm\"a}},\ }\bibfield  {title} {\bibinfo
  {title} {Band geometry, berry curvature, and superfluid weight},\ }\href
  {https://doi.org/10.1103/PhysRevB.95.024515} {\bibfield  {journal} {\bibinfo
  {journal} {Phys. Rev. B}\ }\textbf {\bibinfo {volume} {95}},\ \bibinfo
  {pages} {024515} (\bibinfo {year} {2017})}\BibitemShut {NoStop}%
\bibitem [{\citenamefont {Iskin}(2018)}]{Iskin_2018}%
  \BibitemOpen
  \bibfield  {author} {\bibinfo {author} {\bibfnamefont {M.}~\bibnamefont
  {Iskin}},\ }\bibfield  {title} {\bibinfo {title} {Quantum-metric contribution
  to the pair mass in spin-orbit-coupled fermi superfluids},\ }\href
  {https://doi.org/10.1103/PhysRevA.97.033625} {\bibfield  {journal} {\bibinfo
  {journal} {Phys. Rev. A}\ }\textbf {\bibinfo {volume} {97}},\ \bibinfo
  {pages} {033625} (\bibinfo {year} {2018})}\BibitemShut {NoStop}%
\bibitem [{\citenamefont {Xie}\ \emph {et~al.}(2020)\citenamefont {Xie},
  \citenamefont {Song}, \citenamefont {Lian},\ and\ \citenamefont
  {Bernevig}}]{Xie_2020}%
  \BibitemOpen
  \bibfield  {author} {\bibinfo {author} {\bibfnamefont {F.}~\bibnamefont
  {Xie}}, \bibinfo {author} {\bibfnamefont {Z.}~\bibnamefont {Song}}, \bibinfo
  {author} {\bibfnamefont {B.}~\bibnamefont {Lian}},\ and\ \bibinfo {author}
  {\bibfnamefont {B.~A.}\ \bibnamefont {Bernevig}},\ }\bibfield  {title}
  {\bibinfo {title} {Topology-bounded superfluid weight in twisted bilayer
  graphene},\ }\href {https://doi.org/10.1103/PhysRevLett.124.167002}
  {\bibfield  {journal} {\bibinfo  {journal} {Phys. Rev. Lett.}\ }\textbf
  {\bibinfo {volume} {124}},\ \bibinfo {pages} {167002} (\bibinfo {year}
  {2020})}\BibitemShut {NoStop}%
\bibitem [{\citenamefont {Rossi}(2021)}]{Rossi_2021}%
  \BibitemOpen
  \bibfield  {author} {\bibinfo {author} {\bibfnamefont {E.}~\bibnamefont
  {Rossi}},\ }\bibfield  {title} {\bibinfo {title} {Quantum metric and
  correlated states in two-dimensional systems},\ }\href
  {https://doi.org/https://doi.org/10.1016/j.cossms.2021.100952} {\bibfield
  {journal} {\bibinfo  {journal} {Current Opinion in Solid State and Materials
  Science}\ }\textbf {\bibinfo {volume} {25}},\ \bibinfo {pages} {100952}
  (\bibinfo {year} {2021})}\BibitemShut {NoStop}%
\bibitem [{\citenamefont {Verma}\ \emph {et~al.}(2021)\citenamefont {Verma},
  \citenamefont {Hazra},\ and\ \citenamefont {Randeria}}]{Verma_2021}%
  \BibitemOpen
  \bibfield  {author} {\bibinfo {author} {\bibfnamefont {N.}~\bibnamefont
  {Verma}}, \bibinfo {author} {\bibfnamefont {T.}~\bibnamefont {Hazra}},\ and\
  \bibinfo {author} {\bibfnamefont {M.}~\bibnamefont {Randeria}},\ }\bibfield
  {title} {\bibinfo {title} {Optical spectral weight, phase stiffness, and
  $t_c$ bounds for trivial and topological flat band superconductors},\ }\href
  {https://doi.org/10.1073/pnas.2106744118} {\bibfield  {journal} {\bibinfo
  {journal} {Proceedings of the National Academy of Sciences}\ }\textbf
  {\bibinfo {volume} {118}},\ \bibinfo {pages} {e2106744118} (\bibinfo {year}
  {2021})},\ \Eprint
  {https://arxiv.org/abs/https://www.pnas.org/doi/pdf/10.1073/pnas.2106744118}
  {https://www.pnas.org/doi/pdf/10.1073/pnas.2106744118} \BibitemShut {NoStop}%
\bibitem [{\citenamefont {Ahn}\ and\ \citenamefont {Nagaosa}(2021)}]{Ahn_2021}%
  \BibitemOpen
  \bibfield  {author} {\bibinfo {author} {\bibfnamefont {J.}~\bibnamefont
  {Ahn}}\ and\ \bibinfo {author} {\bibfnamefont {N.}~\bibnamefont {Nagaosa}},\
  }\bibfield  {title} {\bibinfo {title} {Superconductivity-induced spectral
  weight transfer due to quantum geometry},\ }\href
  {https://doi.org/10.1103/PhysRevB.104.L100501} {\bibfield  {journal}
  {\bibinfo  {journal} {Phys. Rev. B}\ }\textbf {\bibinfo {volume} {104}},\
  \bibinfo {pages} {L100501} (\bibinfo {year} {2021})}\BibitemShut {NoStop}%
\bibitem [{\citenamefont {Simon}\ \emph {et~al.}(2022)\citenamefont {Simon},
  \citenamefont {Gabay}, \citenamefont {Goerbig},\ and\ \citenamefont
  {Pagot}}]{Simon_2022}%
  \BibitemOpen
  \bibfield  {author} {\bibinfo {author} {\bibfnamefont {F.}~\bibnamefont
  {Simon}}, \bibinfo {author} {\bibfnamefont {M.}~\bibnamefont {Gabay}},
  \bibinfo {author} {\bibfnamefont {M.~O.}\ \bibnamefont {Goerbig}},\ and\
  \bibinfo {author} {\bibfnamefont {L.}~\bibnamefont {Pagot}},\ }\bibfield
  {title} {\bibinfo {title} {Role of the berry curvature on bcs-type
  superconductivity in two-dimensional materials},\ }\href
  {https://doi.org/10.1103/PhysRevB.106.214512} {\bibfield  {journal} {\bibinfo
   {journal} {Phys. Rev. B}\ }\textbf {\bibinfo {volume} {106}},\ \bibinfo
  {pages} {214512} (\bibinfo {year} {2022})}\BibitemShut {NoStop}%
\bibitem [{\citenamefont {Herzog-Arbeitman}\ \emph {et~al.}(2022)\citenamefont
  {Herzog-Arbeitman}, \citenamefont {Peri}, \citenamefont {Schindler},
  \citenamefont {Huber},\ and\ \citenamefont {Bernevig}}]{Herzog_2022}%
  \BibitemOpen
  \bibfield  {author} {\bibinfo {author} {\bibfnamefont {J.}~\bibnamefont
  {Herzog-Arbeitman}}, \bibinfo {author} {\bibfnamefont {V.}~\bibnamefont
  {Peri}}, \bibinfo {author} {\bibfnamefont {F.}~\bibnamefont {Schindler}},
  \bibinfo {author} {\bibfnamefont {S.~D.}\ \bibnamefont {Huber}},\ and\
  \bibinfo {author} {\bibfnamefont {B.~A.}\ \bibnamefont {Bernevig}},\
  }\bibfield  {title} {\bibinfo {title} {Superfluid weight bounds from symmetry
  and quantum geometry in flat bands},\ }\href
  {https://doi.org/10.1103/PhysRevLett.128.087002} {\bibfield  {journal}
  {\bibinfo  {journal} {Phys. Rev. Lett.}\ }\textbf {\bibinfo {volume} {128}},\
  \bibinfo {pages} {087002} (\bibinfo {year} {2022})}\BibitemShut {NoStop}%
\bibitem [{\citenamefont {Huhtinen}\ \emph {et~al.}(2022)\citenamefont
  {Huhtinen}, \citenamefont {Herzog-Arbeitman}, \citenamefont {Chew},
  \citenamefont {Bernevig},\ and\ \citenamefont {T\"orm\"a}}]{Huhtinen_2022}%
  \BibitemOpen
  \bibfield  {author} {\bibinfo {author} {\bibfnamefont {K.-E.}\ \bibnamefont
  {Huhtinen}}, \bibinfo {author} {\bibfnamefont {J.}~\bibnamefont
  {Herzog-Arbeitman}}, \bibinfo {author} {\bibfnamefont {A.}~\bibnamefont
  {Chew}}, \bibinfo {author} {\bibfnamefont {B.~A.}\ \bibnamefont {Bernevig}},\
  and\ \bibinfo {author} {\bibfnamefont {P.}~\bibnamefont {T\"orm\"a}},\
  }\bibfield  {title} {\bibinfo {title} {Revisiting flat band
  superconductivity: Dependence on minimal quantum metric and band touchings},\
  }\href {https://doi.org/10.1103/PhysRevB.106.014518} {\bibfield  {journal}
  {\bibinfo  {journal} {Phys. Rev. B}\ }\textbf {\bibinfo {volume} {106}},\
  \bibinfo {pages} {014518} (\bibinfo {year} {2022})}\BibitemShut {NoStop}%
\bibitem [{\citenamefont {T{\"o}rm{\"a}}\ \emph {et~al.}(2022)\citenamefont
  {T{\"o}rm{\"a}}, \citenamefont {Peotta},\ and\ \citenamefont
  {Bernevig}}]{Torma_2022}%
  \BibitemOpen
  \bibfield  {author} {\bibinfo {author} {\bibfnamefont {P.}~\bibnamefont
  {T{\"o}rm{\"a}}}, \bibinfo {author} {\bibfnamefont {S.}~\bibnamefont
  {Peotta}},\ and\ \bibinfo {author} {\bibfnamefont {B.~A.}\ \bibnamefont
  {Bernevig}},\ }\bibfield  {title} {\bibinfo {title} {Superconductivity,
  superfluidity and quantum geometry in twisted multilayer systems},\ }\href
  {https://doi.org/10.1038/s42254-022-00466-y} {\bibfield  {journal} {\bibinfo
  {journal} {Nature Reviews Physics}\ }\textbf {\bibinfo {volume} {4}},\
  \bibinfo {pages} {528} (\bibinfo {year} {2022})}\BibitemShut {NoStop}%
\bibitem [{\citenamefont {Peotta}\ \emph {et~al.}()\citenamefont {Peotta},
  \citenamefont {Huhtinen},\ and\ \citenamefont {Torma}}]{Peotta_2023}%
  \BibitemOpen
  \bibfield  {author} {\bibinfo {author} {\bibfnamefont {S.}~\bibnamefont
  {Peotta}}, \bibinfo {author} {\bibfnamefont {K.}~\bibnamefont {Huhtinen}},\
  and\ \bibinfo {author} {\bibfnamefont {P.}~\bibnamefont {Torma}},\ }\bibfield
   {title} {\bibinfo {title} {Quantum geometry in superfluidity and
  superconductivity},\ }\href@noop {} {\bibinfo  {journal} {arXiv:2308.08248}\
  }\BibitemShut {NoStop}%
\bibitem [{\citenamefont {Hu}\ \emph {et~al.}()\citenamefont {Hu},
  \citenamefont {Rossi},\ and\ \citenamefont {Barlas}}]{Hu_2023}%
  \BibitemOpen
\bibfield  {journal} {  }\bibfield  {author} {\bibinfo {author} {\bibfnamefont
  {X.}~\bibnamefont {Hu}}, \bibinfo {author} {\bibfnamefont {E.}~\bibnamefont
  {Rossi}},\ and\ \bibinfo {author} {\bibfnamefont {Y.}~\bibnamefont
  {Barlas}},\ }\bibfield  {title} {\bibinfo {title} {Effect of inversion
  asymmetry on bilayer graphene's superconducting and exciton condensates},\
  }\href@noop {} {\bibinfo  {journal} {arXiv:2304.04825}\ }\BibitemShut
  {NoStop}%
\bibitem [{\citenamefont {Tian}\ \emph {et~al.}(2023)\citenamefont {Tian},
  \citenamefont {Gao}, \citenamefont {Zhang}, \citenamefont {Che},
  \citenamefont {Xu}, \citenamefont {Cheung}, \citenamefont {Watanabe},
  \citenamefont {Taniguchi}, \citenamefont {Randeria}, \citenamefont {Zhang},
  \citenamefont {Lau},\ and\ \citenamefont {Bockrath}}]{Tian_2023}%
  \BibitemOpen
\bibfield  {journal} {  }\bibfield  {author} {\bibinfo {author} {\bibfnamefont
  {H.}~\bibnamefont {Tian}}, \bibinfo {author} {\bibfnamefont {X.}~\bibnamefont
  {Gao}}, \bibinfo {author} {\bibfnamefont {Y.}~\bibnamefont {Zhang}}, \bibinfo
  {author} {\bibfnamefont {S.}~\bibnamefont {Che}}, \bibinfo {author}
  {\bibfnamefont {T.}~\bibnamefont {Xu}}, \bibinfo {author} {\bibfnamefont
  {P.}~\bibnamefont {Cheung}}, \bibinfo {author} {\bibfnamefont
  {K.}~\bibnamefont {Watanabe}}, \bibinfo {author} {\bibfnamefont
  {T.}~\bibnamefont {Taniguchi}}, \bibinfo {author} {\bibfnamefont
  {M.}~\bibnamefont {Randeria}}, \bibinfo {author} {\bibfnamefont
  {F.}~\bibnamefont {Zhang}}, \bibinfo {author} {\bibfnamefont {C.~N.}\
  \bibnamefont {Lau}},\ and\ \bibinfo {author} {\bibfnamefont {M.~W.}\
  \bibnamefont {Bockrath}},\ }\bibfield  {title} {\bibinfo {title} {Evidence
  for dirac flat band superconductivity enabled by quantum geometry},\ }\href
  {https://doi.org/10.1038/s41586-022-05576-2} {\bibfield  {journal} {\bibinfo
  {journal} {Nature}\ }\textbf {\bibinfo {volume} {614}},\ \bibinfo {pages}
  {440} (\bibinfo {year} {2023})}\BibitemShut {NoStop}%
\bibitem [{\citenamefont {Mao}\ and\ \citenamefont
  {Chowdhury}(2023)}]{Mao_2023}%
  \BibitemOpen
  \bibfield  {author} {\bibinfo {author} {\bibfnamefont {D.}~\bibnamefont
  {Mao}}\ and\ \bibinfo {author} {\bibfnamefont {D.}~\bibnamefont
  {Chowdhury}},\ }\bibfield  {title} {\bibinfo {title} {Diamagnetic response
  and phase stiffness for interacting isolated narrow bands},\ }\href
  {https://doi.org/10.1073/pnas.2217816120} {\bibfield  {journal} {\bibinfo
  {journal} {Proceedings of the National Academy of Sciences}\ }\textbf
  {\bibinfo {volume} {120}},\ \bibinfo {pages} {e2217816120} (\bibinfo {year}
  {2023})},\ \Eprint
  {https://arxiv.org/abs/https://www.pnas.org/doi/pdf/10.1073/pnas.2217816120}
  {https://www.pnas.org/doi/pdf/10.1073/pnas.2217816120} \BibitemShut {NoStop}%
\bibitem [{\citenamefont {Verma}\ \emph {et~al.}()\citenamefont {Verma},
  \citenamefont {Guerci},\ and\ \citenamefont {Raquel~Queiroz}}]{Verma_2023}%
  \BibitemOpen
  \bibfield  {author} {\bibinfo {author} {\bibfnamefont {N.}~\bibnamefont
  {Verma}}, \bibinfo {author} {\bibfnamefont {D.}~\bibnamefont {Guerci}},\ and\
  \bibinfo {author} {\bibfnamefont {R.}~\bibnamefont {Raquel~Queiroz}},\
  }\bibfield  {title} {\bibinfo {title} {Geometric stiffness in interlayer
  exciton condensates},\ }\href@noop {} {\bibinfo  {journal}
  {arXiv:2307.01253}\ }\BibitemShut {NoStop}%
\bibitem [{\citenamefont {Xie}\ \emph {et~al.}()\citenamefont {Xie},
  \citenamefont {Ghaemi}, \citenamefont {Mitrano},\ and\ \citenamefont
  {Uchoa}}]{Uchoa_2023}%
  \BibitemOpen
\bibfield  {journal} {  }\bibfield  {author} {\bibinfo {author} {\bibfnamefont
  {H.-Y.}\ \bibnamefont {Xie}}, \bibinfo {author} {\bibfnamefont
  {P.}~\bibnamefont {Ghaemi}}, \bibinfo {author} {\bibfnamefont
  {M.}~\bibnamefont {Mitrano}},\ and\ \bibinfo {author} {\bibfnamefont
  {B.}~\bibnamefont {Uchoa}},\ }\bibfield  {title} {\bibinfo {title} {Theory of
  topological exciton insulators and condensates in flat chern bands},\
  }\href@noop {} {\bibinfo  {journal} {arXiv:2311.04970}\ }\BibitemShut
  {NoStop}%
\bibitem [{\citenamefont {Srivastava}\ and\ \citenamefont
  {Imamo\ifmmode~\breve{g}\else \u{g}\fi{}lu}(2015)}]{Srivastava_2015}%
  \BibitemOpen
\bibfield  {journal} {  }\bibfield  {author} {\bibinfo {author} {\bibfnamefont
  {A.}~\bibnamefont {Srivastava}}\ and\ \bibinfo {author} {\bibfnamefont
  {A.~m.~c.}\ \bibnamefont {Imamo\ifmmode~\breve{g}\else \u{g}\fi{}lu}},\
  }\bibfield  {title} {\bibinfo {title} {Signatures of bloch-band geometry on
  excitons: Nonhydrogenic spectra in transition-metal dichalcogenides},\ }\href
  {https://doi.org/10.1103/PhysRevLett.115.166802} {\bibfield  {journal}
  {\bibinfo  {journal} {Phys. Rev. Lett.}\ }\textbf {\bibinfo {volume} {115}},\
  \bibinfo {pages} {166802} (\bibinfo {year} {2015})}\BibitemShut {NoStop}%
\bibitem [{\citenamefont {Zhou}\ \emph {et~al.}(2015)\citenamefont {Zhou},
  \citenamefont {Shan}, \citenamefont {Yao},\ and\ \citenamefont
  {Xiao}}]{Zhou_2015}%
  \BibitemOpen
  \bibfield  {author} {\bibinfo {author} {\bibfnamefont {J.}~\bibnamefont
  {Zhou}}, \bibinfo {author} {\bibfnamefont {W.-Y.}\ \bibnamefont {Shan}},
  \bibinfo {author} {\bibfnamefont {W.}~\bibnamefont {Yao}},\ and\ \bibinfo
  {author} {\bibfnamefont {D.}~\bibnamefont {Xiao}},\ }\bibfield  {title}
  {\bibinfo {title} {Berry phase modification to the energy spectrum of
  excitons},\ }\href {https://doi.org/10.1103/PhysRevLett.115.166803}
  {\bibfield  {journal} {\bibinfo  {journal} {Phys. Rev. Lett.}\ }\textbf
  {\bibinfo {volume} {115}},\ \bibinfo {pages} {166803} (\bibinfo {year}
  {2015})}\BibitemShut {NoStop}%
\bibitem [{\citenamefont {Song}\ and\ \citenamefont
  {Rudner}(2016)}]{Song_2016}%
  \BibitemOpen
  \bibfield  {author} {\bibinfo {author} {\bibfnamefont {J.~C.~W.}\
  \bibnamefont {Song}}\ and\ \bibinfo {author} {\bibfnamefont {M.~S.}\
  \bibnamefont {Rudner}},\ }\bibfield  {title} {\bibinfo {title} {Chiral
  plasmons without magnetic field},\ }\href
  {https://doi.org/10.1073/pnas.1519086113} {\bibfield  {journal} {\bibinfo
  {journal} {Proceedings of the National Academy of Sciences}\ }\textbf
  {\bibinfo {volume} {113}},\ \bibinfo {pages} {4658} (\bibinfo {year}
  {2016})},\ \Eprint
  {https://arxiv.org/abs/https://www.pnas.org/doi/pdf/10.1073/pnas.1519086113}
  {https://www.pnas.org/doi/pdf/10.1073/pnas.1519086113} \BibitemShut {NoStop}%
\bibitem [{\citenamefont {Arora}\ \emph {et~al.}(2022)\citenamefont {Arora},
  \citenamefont {Rudner},\ and\ \citenamefont {Song}}]{Arora_2022}%
  \BibitemOpen
  \bibfield  {author} {\bibinfo {author} {\bibfnamefont {A.}~\bibnamefont
  {Arora}}, \bibinfo {author} {\bibfnamefont {M.~S.}\ \bibnamefont {Rudner}},\
  and\ \bibinfo {author} {\bibfnamefont {J.~C.~W.}\ \bibnamefont {Song}},\
  }\bibfield  {title} {\bibinfo {title} {Quantum plasmonic nonreciprocity in
  parity-violating magnets},\ }\href
  {https://doi.org/10.1021/acs.nanolett.2c03126} {\bibfield  {journal}
  {\bibinfo  {journal} {Nano Letters}\ }\textbf {\bibinfo {volume} {22}},\
  \bibinfo {pages} {9351} (\bibinfo {year} {2022})}\BibitemShut {NoStop}%
\bibitem [{\citenamefont {Yao}\ and\ \citenamefont {Niu}(2008)}]{Yao_2008}%
  \BibitemOpen
  \bibfield  {author} {\bibinfo {author} {\bibfnamefont {W.}~\bibnamefont
  {Yao}}\ and\ \bibinfo {author} {\bibfnamefont {Q.}~\bibnamefont {Niu}},\
  }\bibfield  {title} {\bibinfo {title} {Berry phase effect on the exciton
  transport and on the exciton bose-einstein condensate},\ }\href
  {https://doi.org/10.1103/PhysRevLett.101.106401} {\bibfield  {journal}
  {\bibinfo  {journal} {Phys. Rev. Lett.}\ }\textbf {\bibinfo {volume} {101}},\
  \bibinfo {pages} {106401} (\bibinfo {year} {2008})}\BibitemShut {NoStop}%
\bibitem [{\citenamefont {Kwan}\ \emph {et~al.}(2021)\citenamefont {Kwan},
  \citenamefont {Hu}, \citenamefont {Simon},\ and\ \citenamefont
  {Parameswaran}}]{Kwan_2021}%
  \BibitemOpen
  \bibfield  {author} {\bibinfo {author} {\bibfnamefont {Y.~H.}\ \bibnamefont
  {Kwan}}, \bibinfo {author} {\bibfnamefont {Y.}~\bibnamefont {Hu}}, \bibinfo
  {author} {\bibfnamefont {S.~H.}\ \bibnamefont {Simon}},\ and\ \bibinfo
  {author} {\bibfnamefont {S.~A.}\ \bibnamefont {Parameswaran}},\ }\bibfield
  {title} {\bibinfo {title} {Exciton band topology in spontaneous quantum
  anomalous hall insulators: Applications to twisted bilayer graphene},\ }\href
  {https://doi.org/10.1103/PhysRevLett.126.137601} {\bibfield  {journal}
  {\bibinfo  {journal} {Phys. Rev. Lett.}\ }\textbf {\bibinfo {volume} {126}},\
  \bibinfo {pages} {137601} (\bibinfo {year} {2021})}\BibitemShut {NoStop}%
\bibitem [{\citenamefont {Cao}\ \emph {et~al.}(2021{\natexlab{a}})\citenamefont
  {Cao}, \citenamefont {Fertig},\ and\ \citenamefont {Brey}}]{Cao_PRB_2021}%
  \BibitemOpen
  \bibfield  {author} {\bibinfo {author} {\bibfnamefont {J.}~\bibnamefont
  {Cao}}, \bibinfo {author} {\bibfnamefont {H.~A.}\ \bibnamefont {Fertig}},\
  and\ \bibinfo {author} {\bibfnamefont {L.}~\bibnamefont {Brey}},\ }\bibfield
  {title} {\bibinfo {title} {Quantum geometric exciton drift velocity},\ }\href
  {https://doi.org/10.1103/PhysRevB.103.115422} {\bibfield  {journal} {\bibinfo
   {journal} {Phys. Rev. B}\ }\textbf {\bibinfo {volume} {103}},\ \bibinfo
  {pages} {115422} (\bibinfo {year} {2021}{\natexlab{a}})}\BibitemShut
  {NoStop}%
\bibitem [{\citenamefont {Tang}\ \emph {et~al.}()\citenamefont {Tang},
  \citenamefont {Wang},\ and\ \citenamefont {Yu}}]{Tang_2023}%
  \BibitemOpen
  \bibfield  {author} {\bibinfo {author} {\bibfnamefont {J.}~\bibnamefont
  {Tang}}, \bibinfo {author} {\bibfnamefont {S.}~\bibnamefont {Wang}},\ and\
  \bibinfo {author} {\bibfnamefont {H.}~\bibnamefont {Yu}},\ }\bibfield
  {title} {\bibinfo {title} {Inheritance of the exciton geometric structure
  from bloch electrons in two-dimensional layered semiconductors},\ }\href@noop
  {} {\bibinfo  {journal} {arXiv:2310.14856}\ }\BibitemShut {NoStop}%
\bibitem [{\citenamefont {Chaudhary}\ \emph {et~al.}(2021)\citenamefont
  {Chaudhary}, \citenamefont {Knapp},\ and\ \citenamefont
  {Refael}}]{Chaudhary_2021}%
  \BibitemOpen
\bibfield  {journal} {  }\bibfield  {author} {\bibinfo {author} {\bibfnamefont
  {S.}~\bibnamefont {Chaudhary}}, \bibinfo {author} {\bibfnamefont
  {C.}~\bibnamefont {Knapp}},\ and\ \bibinfo {author} {\bibfnamefont
  {G.}~\bibnamefont {Refael}},\ }\bibfield  {title} {\bibinfo {title}
  {Anomalous exciton transport in response to a uniform in-plane electric
  field},\ }\href {https://doi.org/10.1103/PhysRevB.103.165119} {\bibfield
  {journal} {\bibinfo  {journal} {Phys. Rev. B}\ }\textbf {\bibinfo {volume}
  {103}},\ \bibinfo {pages} {165119} (\bibinfo {year} {2021})}\BibitemShut
  {NoStop}%
\bibitem [{\citenamefont {Sawada}\ \emph {et~al.}(1957)\citenamefont {Sawada},
  \citenamefont {Brueckner}, \citenamefont {Fukuda},\ and\ \citenamefont
  {Brout}}]{Sawada:1957aa}%
  \BibitemOpen
  \bibfield  {author} {\bibinfo {author} {\bibfnamefont {K.}~\bibnamefont
  {Sawada}}, \bibinfo {author} {\bibfnamefont {K.~A.}\ \bibnamefont
  {Brueckner}}, \bibinfo {author} {\bibfnamefont {N.}~\bibnamefont {Fukuda}},\
  and\ \bibinfo {author} {\bibfnamefont {R.}~\bibnamefont {Brout}},\ }\bibfield
   {title} {\bibinfo {title} {Correlation energy of an electron gas at high
  density: Plasma oscillations},\ }\href
  {https://doi.org/10.1103/PhysRev.108.507} {\bibfield  {journal} {\bibinfo
  {journal} {Physical Review}\ }\textbf {\bibinfo {volume} {108}},\ \bibinfo
  {pages} {507} (\bibinfo {year} {1957})}\BibitemShut {NoStop}%
\bibitem [{\citenamefont {Cao}\ \emph {et~al.}(2021{\natexlab{b}})\citenamefont
  {Cao}, \citenamefont {Fertig},\ and\ \citenamefont {Brey}}]{Cao_PRL_2021}%
  \BibitemOpen
  \bibfield  {author} {\bibinfo {author} {\bibfnamefont {J.}~\bibnamefont
  {Cao}}, \bibinfo {author} {\bibfnamefont {H.~A.}\ \bibnamefont {Fertig}},\
  and\ \bibinfo {author} {\bibfnamefont {L.}~\bibnamefont {Brey}},\ }\bibfield
  {title} {\bibinfo {title} {Quantum internal structure of plasmons},\ }\href
  {https://doi.org/10.1103/PhysRevLett.127.196403} {\bibfield  {journal}
  {\bibinfo  {journal} {Phys. Rev. Lett.}\ }\textbf {\bibinfo {volume} {127}},\
  \bibinfo {pages} {196403} (\bibinfo {year} {2021}{\natexlab{b}})}\BibitemShut
  {NoStop}%
\bibitem [{\citenamefont {Cao}\ \emph {et~al.}(2022)\citenamefont {Cao},
  \citenamefont {Fertig},\ and\ \citenamefont {Brey}}]{Cao_2022}%
  \BibitemOpen
  \bibfield  {author} {\bibinfo {author} {\bibfnamefont {J.}~\bibnamefont
  {Cao}}, \bibinfo {author} {\bibfnamefont {H.~A.}\ \bibnamefont {Fertig}},\
  and\ \bibinfo {author} {\bibfnamefont {L.}~\bibnamefont {Brey}},\ }\bibfield
  {title} {\bibinfo {title} {Plasmonic transverse dipole moment in chiral
  fermion nanowires},\ }\href {https://doi.org/10.1103/PhysRevB.106.165125}
  {\bibfield  {journal} {\bibinfo  {journal} {Phys. Rev. B}\ }\textbf {\bibinfo
  {volume} {106}},\ \bibinfo {pages} {165125} (\bibinfo {year}
  {2022})}\BibitemShut {NoStop}%
\bibitem [{\citenamefont {Ulloa}\ \emph {et~al.}(2002)\citenamefont {Ulloa},
  \citenamefont {Govorov}, \citenamefont {Kalameitsev}, \citenamefont
  {Warburton},\ and\ \citenamefont {Karrai}}]{Ulloa_2002}%
  \BibitemOpen
  \bibfield  {author} {\bibinfo {author} {\bibfnamefont {S.}~\bibnamefont
  {Ulloa}}, \bibinfo {author} {\bibfnamefont {A.}~\bibnamefont {Govorov}},
  \bibinfo {author} {\bibfnamefont {A.}~\bibnamefont {Kalameitsev}}, \bibinfo
  {author} {\bibfnamefont {R.}~\bibnamefont {Warburton}},\ and\ \bibinfo
  {author} {\bibfnamefont {K.}~\bibnamefont {Karrai}},\ }\bibfield  {title}
  {\bibinfo {title} {Magnetoexcitons in quantum-ring structures: a novel
  magnetic interference effect},\ }\href@noop {} {\bibfield  {journal}
  {\bibinfo  {journal} {Physica E: Low-dimensional Systems and Nanostructures}\
  }\textbf {\bibinfo {volume} {12}},\ \bibinfo {pages} {790} (\bibinfo {year}
  {2002})}\BibitemShut {NoStop}%
\bibitem [{\citenamefont {Vanderbilt}(2018)}]{Vanderbilt_book}%
  \BibitemOpen
  \bibfield  {author} {\bibinfo {author} {\bibfnamefont {D.}~\bibnamefont
  {Vanderbilt}},\ }\href@noop {} {\emph {\bibinfo {title} {Berry Phases in
  Electronic Structure Theory}}}\ (\bibinfo  {publisher} {Cambridge University
  Press},\ \bibinfo {year} {2018})\BibitemShut {NoStop}%
\bibitem [{\citenamefont {Yoshioka}(2002)}]{Yoshioka_book}%
  \BibitemOpen
  \bibfield  {author} {\bibinfo {author} {\bibfnamefont {D.}~\bibnamefont
  {Yoshioka}},\ }\href@noop {} {\emph {\bibinfo {title} {The Quantum Hall
  Effect}}}\ (\bibinfo  {publisher} {Springer},\ \bibinfo {year}
  {2002})\BibitemShut {NoStop}%
\bibitem [{\citenamefont {Lerner}\ and\ \citenamefont
  {YE~Lozovik}(1979)}]{Lerner_1979}%
  \BibitemOpen
  \bibfield  {author} {\bibinfo {author} {\bibfnamefont {I.}~\bibnamefont
  {Lerner}}\ and\ \bibinfo {author} {\bibfnamefont {Y.}~\bibnamefont
  {YE~Lozovik}},\ }\href@noop {} {\bibfield  {journal} {\bibinfo  {journal}
  {Journal of Physics C Solid State Physics}\ }\textbf {\bibinfo {volume}
  {12}},\ \bibinfo {pages} {L501} (\bibinfo {year} {1979})}\BibitemShut
  {NoStop}%
\bibitem [{\citenamefont {{Bychkov}}\ \emph {et~al.}(1981)\citenamefont
  {{Bychkov}}, \citenamefont {{Iordanski{\v{i}}}},\ and\ \citenamefont
  {{{\'E}liashberg}}}]{Bychkov_1981}%
  \BibitemOpen
  \bibfield  {author} {\bibinfo {author} {\bibfnamefont {Y.~A.}\ \bibnamefont
  {{Bychkov}}}, \bibinfo {author} {\bibfnamefont {S.~V.}\ \bibnamefont
  {{Iordanski{\v{i}}}}},\ and\ \bibinfo {author} {\bibfnamefont {G.~M.}\
  \bibnamefont {{{\'E}liashberg}}},\ }\bibfield  {title} {\bibinfo {title}
  {{Two-dimensional electrons in a strong magnetic field}},\ }\href@noop {}
  {\bibfield  {journal} {\bibinfo  {journal} {Soviet Journal of Experimental
  and Theoretical Physics Letters}\ }\textbf {\bibinfo {volume} {33}},\
  \bibinfo {pages} {143} (\bibinfo {year} {1981})}\BibitemShut {NoStop}%
\bibitem [{\citenamefont {Kallin}\ and\ \citenamefont
  {Halperin}(1984)}]{Kallin_1984}%
  \BibitemOpen
  \bibfield  {author} {\bibinfo {author} {\bibfnamefont {C.}~\bibnamefont
  {Kallin}}\ and\ \bibinfo {author} {\bibfnamefont {B.~I.}\ \bibnamefont
  {Halperin}},\ }\bibfield  {title} {\bibinfo {title} {Excitations from a
  filled landau level in the two-dimensional electron gas},\ }\href
  {https://doi.org/10.1103/PhysRevB.30.5655} {\bibfield  {journal} {\bibinfo
  {journal} {Phys. Rev. B}\ }\textbf {\bibinfo {volume} {30}},\ \bibinfo
  {pages} {5655} (\bibinfo {year} {1984})}\BibitemShut {NoStop}%
\bibitem [{\citenamefont {Feynman}(1972)}]{Feynman-Book}%
  \BibitemOpen
  \bibfield  {author} {\bibinfo {author} {\bibfnamefont {R.}~\bibnamefont
  {Feynman}},\ }\href@noop {} {\emph {\bibinfo {title} {Statistical
  Mechanics}}}\ (\bibinfo  {publisher} {Benjamin, Reading, MA},\ \bibinfo
  {year} {1972})\BibitemShut {NoStop}%
\bibitem [{\citenamefont {Girvin}\ \emph {et~al.}(1985)\citenamefont {Girvin},
  \citenamefont {MacDonald},\ and\ \citenamefont {Platzman}}]{Girvin_1985}%
  \BibitemOpen
  \bibfield  {author} {\bibinfo {author} {\bibfnamefont {S.~M.}\ \bibnamefont
  {Girvin}}, \bibinfo {author} {\bibfnamefont {A.~H.}\ \bibnamefont
  {MacDonald}},\ and\ \bibinfo {author} {\bibfnamefont {P.~M.}\ \bibnamefont
  {Platzman}},\ }\bibfield  {title} {\bibinfo {title} {Collective-excitation
  gap in the fractional quantum hall effect},\ }\href
  {https://doi.org/10.1103/PhysRevLett.54.581} {\bibfield  {journal} {\bibinfo
  {journal} {Phys. Rev. Lett.}\ }\textbf {\bibinfo {volume} {54}},\ \bibinfo
  {pages} {581} (\bibinfo {year} {1985})}\BibitemShut {NoStop}%
\bibitem [{\citenamefont {Girvin}\ \emph {et~al.}(1986)\citenamefont {Girvin},
  \citenamefont {MacDonald},\ and\ \citenamefont {Platzman}}]{Girvin_1986}%
  \BibitemOpen
  \bibfield  {author} {\bibinfo {author} {\bibfnamefont {S.~M.}\ \bibnamefont
  {Girvin}}, \bibinfo {author} {\bibfnamefont {A.~H.}\ \bibnamefont
  {MacDonald}},\ and\ \bibinfo {author} {\bibfnamefont {P.~M.}\ \bibnamefont
  {Platzman}},\ }\bibfield  {title} {\bibinfo {title} {Magneto-roton theory of
  collective excitations in the fractional quantum hall effect},\ }\href
  {https://doi.org/10.1103/PhysRevB.33.2481} {\bibfield  {journal} {\bibinfo
  {journal} {Phys. Rev. B}\ }\textbf {\bibinfo {volume} {33}},\ \bibinfo
  {pages} {2481} (\bibinfo {year} {1986})}\BibitemShut {NoStop}%
\bibitem [{\citenamefont {Tsui}\ \emph {et~al.}(1982)\citenamefont {Tsui},
  \citenamefont {Stormer},\ and\ \citenamefont {Gossard}}]{Tsui_1982}%
  \BibitemOpen
  \bibfield  {author} {\bibinfo {author} {\bibfnamefont {D.~C.}\ \bibnamefont
  {Tsui}}, \bibinfo {author} {\bibfnamefont {H.~L.}\ \bibnamefont {Stormer}},\
  and\ \bibinfo {author} {\bibfnamefont {A.~C.}\ \bibnamefont {Gossard}},\
  }\bibfield  {title} {\bibinfo {title} {Two-dimensional magnetotransport in
  the extreme quantum limit},\ }\href
  {https://doi.org/10.1103/PhysRevLett.48.1559} {\bibfield  {journal} {\bibinfo
   {journal} {Phys. Rev. Lett.}\ }\textbf {\bibinfo {volume} {48}},\ \bibinfo
  {pages} {1559} (\bibinfo {year} {1982})}\BibitemShut {NoStop}%
\bibitem [{\citenamefont {Laughlin}(1983)}]{Laughlin_1983}%
  \BibitemOpen
  \bibfield  {author} {\bibinfo {author} {\bibfnamefont {R.~B.}\ \bibnamefont
  {Laughlin}},\ }\bibfield  {title} {\bibinfo {title} {Anomalous quantum hall
  effect: An incompressible quantum fluid with fractionally charged
  excitations},\ }\href {https://doi.org/10.1103/PhysRevLett.50.1395}
  {\bibfield  {journal} {\bibinfo  {journal} {Phys. Rev. Lett.}\ }\textbf
  {\bibinfo {volume} {50}},\ \bibinfo {pages} {1395} (\bibinfo {year}
  {1983})}\BibitemShut {NoStop}%
\bibitem [{\citenamefont {Haldane}\ and\ \citenamefont
  {Rezayi}(1985)}]{Haldane_1985}%
  \BibitemOpen
  \bibfield  {author} {\bibinfo {author} {\bibfnamefont {F.~D.~M.}\
  \bibnamefont {Haldane}}\ and\ \bibinfo {author} {\bibfnamefont {E.~H.}\
  \bibnamefont {Rezayi}},\ }\bibfield  {title} {\bibinfo {title} {Periodic
  laughlin-jastrow wave functions for the fractional quantized hall effect},\
  }\href {https://doi.org/10.1103/PhysRevB.31.2529} {\bibfield  {journal}
  {\bibinfo  {journal} {Phys. Rev. B}\ }\textbf {\bibinfo {volume} {31}},\
  \bibinfo {pages} {2529} (\bibinfo {year} {1985})}\BibitemShut {NoStop}%
\bibitem [{\citenamefont {Jain}(2007)}]{Jain_book}%
  \BibitemOpen
  \bibfield  {author} {\bibinfo {author} {\bibfnamefont {J.}~\bibnamefont
  {Jain}},\ }\href@noop {} {\emph {\bibinfo {title} {Composite Fermions}}}\
  (\bibinfo  {publisher} {Cambridge University Press},\ \bibinfo {year}
  {2007})\BibitemShut {NoStop}%
\bibitem [{\citenamefont {Griffiths}\ \emph {et~al.}(2022)\citenamefont
  {Griffiths}, \citenamefont {Derbes}, \citenamefont {Sohn},\ and\
  \citenamefont {Eds.}}]{Coleman_book}%
  \BibitemOpen
  \bibfield  {author} {\bibinfo {author} {\bibfnamefont {D.}~\bibnamefont
  {Griffiths}}, \bibinfo {author} {\bibfnamefont {D.}~\bibnamefont {Derbes}},
  \bibinfo {author} {\bibfnamefont {R.}~\bibnamefont {Sohn}},\ and\ \bibinfo
  {author} {\bibnamefont {Eds.}},\ }\href@noop {} {\emph {\bibinfo {title}
  {Sidney Coleman's Lectures onf Relativity}}}\ (\bibinfo  {publisher}
  {Cambridge University Press},\ \bibinfo {year} {2022})\BibitemShut {NoStop}%
\bibitem [{com()}]{comment}%
  \BibitemOpen
  \href@noop {} {}\bibinfo {note} {Due to periodic boundary conditions, there
  will also be jumps in ${\bf r}$ when a particle ``exits'' one boundary and
  enters another that has been ``stitched'' to it. Provided the resulting lines
  of discontinuity stay on unit cell edges, the value of the QGD will be
  unaffected by the precise choice of path of these discontinuities. This is
  because the total fermion charge in each unit cell must be the same for both
  the physical state and the reference state.}\BibitemShut {Stop}%
\bibitem [{\citenamefont {Repellin}\ \emph {et~al.}(2014)\citenamefont
  {Repellin}, \citenamefont {Neupert}, \citenamefont
  {Papi\ifmmode~\acute{c}\else \'{c}\fi{}},\ and\ \citenamefont
  {Regnault}}]{Repellin_2014}%
  \BibitemOpen
  \bibfield  {author} {\bibinfo {author} {\bibfnamefont {C.}~\bibnamefont
  {Repellin}}, \bibinfo {author} {\bibfnamefont {T.}~\bibnamefont {Neupert}},
  \bibinfo {author} {\bibfnamefont {Z.}~\bibnamefont
  {Papi\ifmmode~\acute{c}\else \'{c}\fi{}}},\ and\ \bibinfo {author}
  {\bibfnamefont {N.}~\bibnamefont {Regnault}},\ }\bibfield  {title} {\bibinfo
  {title} {Single-mode approximation for fractional chern insulators and the
  fractional quantum hall effect on the torus},\ }\href
  {https://doi.org/10.1103/PhysRevB.90.045114} {\bibfield  {journal} {\bibinfo
  {journal} {Phys. Rev. B}\ }\textbf {\bibinfo {volume} {90}},\ \bibinfo
  {pages} {045114} (\bibinfo {year} {2014})}\BibitemShut {NoStop}%
\bibitem [{\citenamefont {Girvin}\ and\ \citenamefont
  {Jach}(1984)}]{Girvin_1984}%
  \BibitemOpen
  \bibfield  {author} {\bibinfo {author} {\bibfnamefont {S.~M.}\ \bibnamefont
  {Girvin}}\ and\ \bibinfo {author} {\bibfnamefont {T.}~\bibnamefont {Jach}},\
  }\bibfield  {title} {\bibinfo {title} {Formalism for the quantum hall effect:
  Hilbert space of analytic functions},\ }\href
  {https://doi.org/10.1103/PhysRevB.29.5617} {\bibfield  {journal} {\bibinfo
  {journal} {Phys. Rev. B}\ }\textbf {\bibinfo {volume} {29}},\ \bibinfo
  {pages} {5617} (\bibinfo {year} {1984})}\BibitemShut {NoStop}%
\bibitem [{HAF()}]{HAF_LB_unpub}%
  \BibitemOpen
  \href@noop {} {}\bibinfo {note} {H.A. Fertig and L. Brey,
  unpublished.}\BibitemShut {Stop}%
\end{thebibliography}
\end{document}